\documentstyle[11pt,epic,eepic]{article}
\title{Scoping Constructs in Logic Programming:\\
Implementation Problems and their Solution}

\author{Gopalan Nadathur\\ Bharat Jayaraman\\ Keehang Kwon}

\newtheorem{proposition}{Proposition}[section]

\def\thebibliography#1{\section*{REFERENCES}\list
 {[\arabic{enumi}]}{\settowidth\labelwidth{[#1]}\leftmargin\labelwidth
 \advance\leftmargin\labelsep
 \usecounter{enumi}}
 \def\newblock{\hskip .11em plus .33em minus .07em}
 \sloppy\clubpenalty4000\widowpenalty4000
 \sfcode`\.=1000\relax}

\makeatletter
\long\def\@makemyfntext#1{ #1}

\long\def\@myfootnotetext#1{\insert\footins{\footnotesize
    \interlinepenalty\interfootnotelinepenalty 
    \splittopskip\footnotesep
    \splitmaxdepth \dp\strutbox \floatingpenalty \@MM
    \hsize\columnwidth \@parboxrestore
   \edef\@currentlabel{\csname p@footnote\endcsname\@thefnmark}\@makemyfntext
    {\rule{\z@}{\footnotesep}\ignorespaces
      #1\strut}}}

\def\myfootnotetext{\@ifnextchar
     [{\@xfootnotenext}{\xdef\@thefnmark{\thempfn}\@myfootnotetext}}
\makeatother

\long\def\ignore#1{}

\newenvironment{numberedlist}
{\begin{list}{\makebox[20pt]{\hss(\arabic{itemno})\enspace}}
             {\usecounter{itemno}\labelwidth 20pt}}{\end{list}}

\newenvironment{alphabetlist}
{\begin{list}{\makebox[20pt]{\hss(\alph{itemno1})\enspace}}
             {\usecounter{itemno1}\labelwidth 20pt}}{\end{list}}

\newenvironment{romanlist}
{\begin{list}{\makebox[20pt]{\hss(\roman{itemno2})\enspace}}
             {\usecounter{itemno2}\labelwidth 20pt}}{\end{list}}

\newenvironment{exmple}{
 \begingroup \begin{tabbing} \hspace{2em}\= \hspace{5em}\= \hspace{5em}\=
\hspace{5em}\= \kill}{
 \end{tabbing}\endgroup}

\newcounter{itemno}

\newcounter{itemno1}

\newcounter{itemno2}

\newcommand{\subs}[3]{[#1/#2]#3}

\newcommand{\ra}{\rightarrow}
\newcommand{\app}{{\ }}
\newcommand{\sep}{\;\vert\;}
\newcommand{\all}{\forall}
\newcommand{\lambdax}[1]{\lambda #1 \,}
\newcommand{\allx}[1]{\forall #1 \,}
\newcommand{\some}{\exists}
\newcommand{\somex}[1]{\exists #1 \,}
\newcommand{\Pscr}{{\cal P}}
\newcommand{\Gscr}{{\cal G}}
\renewcommand{\neg}{\mathord\sim}
\newcommand{\dset}{{D\!s}}

\newcommand{\ie}{{\em i.e.}}

\newcommand{\eg}{{\em e.g.}}

\begin{document}

\bibliographystyle{alpha}

\begin{center}
{\Large {\bf  Scoping Constructs in Logic Programming:\\[10pt]
Implementation Problems and their Solution}}
\\[20pt]
{\bf Gopalan Nadathur}\\
Department of Computer Science\\
University of Chicago\\
Ryerson Laboratory, 1100 E 58th Street\\
Chicago, IL 60637
gopalan@cs.uchicago.edu\\
[10pt]
{\bf Bharat Jayaraman}\\
Department of Computer Science\\
State University of New York at Buffalo\\
Buffalo, NY 14260\\
bharat@cs.buffalo.edu\\
[10pt]
{\bf Keehang Kwon}\\
Department of Computer Science\\
Duke University, Durham, NC 27706\\
kwon@cs.duke.edu
\end{center}

\begin{abstract}
The inclusion of universal quantification and a form of implication in
goals in logic programming is considered.  These additions provide a
logical basis for scoping but they also raise new
implementation problems.  When universal and existential quantifiers
are permitted to appear in mixed order in goals, the devices of logic
variables and unification that are employed in solving existential
goals must be modified to ensure that constraints arising out of the
order of quantification are respected. Suitable modifications that are
based on attaching numerical tags to constants and variables and on
using these tags in unification are described.  The
resulting devices are amenable to an efficient implementation and can,
in fact, be assimilated easily into the usual machinery of the Warren
Abstract Machine (WAM). The provision of implications in goals results
in the possibility of program clauses being added to the program for
the purpose of solving specific subgoals.  A naive scheme based on 
asserting and retracting program clauses does not suffice for
implementing such additions for two reasons.  First, it is necessary
to also support the resurrection of an earlier existing program in the
face of backtracking.  Second, the possibility for implication goals
to be surrounded by quantifiers requires a consideration of the
parameterization of program clauses by bindings for their free
variables. Devices for supporting these additional requirements are
described as also is the integration of these devices into the WAM.
Further extensions to the machine are outlined for handling
higher-order additions to the language. The ideas presented here are
relevant to the implementation of the higher-order logic programming
language $\lambda$Prolog.
\end{abstract}

\section{INTRODUCTION}\label{sec:intro}

This paper examines techniques relevant to the implementation of the
logic programming language $\lambda$Prolog \cite{NM88}. The basis for
this language is provided by a polymorphic version of the logic of
higher-order hereditary Harrop or {\em hohh} formulas \cite{MNPS91}.
At a qualitative level, the logic of {\em hohh} 
formulas represents an amalgamation of extensions to Horn clause logic 
in two different directions.  The extension in one direction is
obtained by including higher-order features --- in the form of
quantification over function and some occurrences of predicate
variables and the replacement of first-order terms by simply typed
lambda terms --- within Horn clauses, thereby producing the logic of
{\it higher-order} Horn clauses \cite{NM90}.  Along the other
direction, Horn clause logic is enhanced 
by permitting universal quantifiers and restricted uses of
implications, resulting in a first-order version of the logic of
hereditary Harrop 
formulas \cite{Miller87lmps,MNPS91}. The combination of these two
logics produces a simply typed version of the logic of {\em hohh}
formulas.  The typing paradigm incorporated in this logic is somewhat
constraining from the perspective of programming. However, it can be
relaxed through the introduction of polymorphism.  The resulting logic
is what constitutes the basis for $\lambda$Prolog.

The enrichments to Horn clause logic that are embodied in the logic
underlying $\lambda$Prolog provide for new features at a programming
level. $\lambda$Prolog is, in fact, a language that manifests these
features and consequently has several novel capabilities in comparison
with a language like Prolog. The usefulness of these capabilities has
lead to a significant interest in the language and systems have been
developed that implement $\lambda$Prolog or a close relative of it
\cite{BR91,EP91,NM88}.  These systems notwithstanding, there has been
little discussion of techniques that are well-suited to the
implementation of such a language.\footnote{There is, however, a
discussion of the implementation problems in \cite{EP91} and also a
systematic development of an interpreter for $\lambda$Prolog within a
functional programming language.}\ The considerations in this paper
are part of an effort that focuses on precisely this issue, with the
ultimate goal of providing an efficient and robust implementation for
$\lambda$Prolog.  We have found the hierarchy of logics described
above a useful structuring device in this endeavor. In particular, we
have been developing an implementation scheme for the full language by
starting with the Warren Abstract Machine (WAM) \cite{War83}, which is
usually employed for Prolog, and considering independently the new
devices that are required for dealing with higher-order aspects,
types, and implications and universal quantifiers.  There is good
reason for adopting such an approach: Unification and backtracking are
central to the implementation of all the logics in question, and
the WAM provides a good framework for an efficient treatment
of these aspects.  Furthermore, the new features in $\lambda$Prolog
are in a sense orthogonal to each other. Consequently there is little
interference between the mechanisms developed for realizing each of
these features, and, in fact, they blend together well in
an overall machine.

In keeping with the above strategy, this paper discusses
implementation methods for one of the new aspects of $\lambda$Prolog,
namely, the provision of implications and universal quantifiers in
goals. It complements, in this respect, other work that we have done
concerning the treatment of higher-order aspects
\cite{Nad94a,NJW92,NW93a} and types \cite{KNW92}. The
particular enrichment considered here is also of interest in its own
right: permitting implications and universal quantifiers in goals
provides the basis for scoping constructs in a language such as
Prolog.  From the perspective of an implementation, the inclusion of
these symbols gives rise to two new kinds of problems. The first kind
of problem arises from the possibility of alternating sequences of
universal and existential quantifiers appearing in goals. Solving a
universally quantified goal requires the introduction of a ``new''
constant. The usual implementation technique employed for an
existential quantifier is to instantiate it with a ``logic'' variable
whose value is determined at a later stage through unification. Care
must be exercised in combining these two strategies.  In order to
guarantee the newness of a constant introduced for a universal
quantifier, this constant must not be allowed to appear in the term
that ultimately instantiates a surrounding existential quantifier. A
proper treatment of unification is necessary for this purpose. The
second kind of problem is caused by the fact that implications in
goals require sets of program clauses to be periodically added and
removed from the program.  While it might seem that a simple-minded
stack-based scheme can be used to implement programs that change in
this manner, there are some complications: First, the program clauses
that may need to be assumed may be ``parameterized'' by bindings for
the free variables occurring in them, requiring them to be treated as
{\it closures}.  Second, backtracking action may require
the reinstatement of a program in existence at some earlier point and
a bookkeeping scheme that makes it possible to carry out this action
in an efficient manner is needed.

In the rest of this paper, we discuss in detail the provision of
implications and universal quantifiers in goals and the new
implementation problems that arise from this enhancement.  This
discussion is structured as follows.  In the next section we present
informally a language that extends Prolog in the manner mentioned and
we illustrate the usefulness of the new features of this language. We
describe a first-order version of this language formally in 
Section \ref{sec:fohh} and discuss the implementation problems that
arise in its context. We then devote our attention to methods for
dealing with these problems.  In Section \ref{sec:interp} we present
an abstract interpreter for our extended language that contains within
it a conceptual scheme for handling universal quantifiers.
This interpreter is naive in its treatment of implication
goals, and the next two sections focus on this issue. In Section
\ref{sec:impl} we present solutions to the two main problems that
arise in this context: the parameterization of program clauses and the
need to resurrect old program contexts on backtracking. An efficient
realization of these solutions within a WAM-like framework is
discussed in Section \ref{sec:rep}.  In Section~\ref{sec:compile}, we
examine the possibility of compilation within our implementation
scheme.  This discussion provides a complete picture of our
implementation ideas and also illustrates the graceful manner in which
the additional machinery fits into that of the WAM.  Although the most
interesting motivating examples for the inclusion of implications and
universal quantifiers in goals involve the use of a higher-order
language, simplicity of exposition dictates that we present our
implementation ideas in a first-order context.  We amend this
situation in Section~\ref{sec:higherorder} by indicating how these
ideas translate to the higher-order context and by describing methods
for dealing with additional aspects of scoping in this context that
are not covered by them. We conclude this paper in Section
\ref{sec:conclusion}.

\section{USES OF IMPLICATIONS AND UNIVERSAL QUANTIFIERS IN
GOALS}\label{sec:examples} 

We shall describe precisely the idea of a ``goal'' in Section
\ref{sec:fohh}; for the moment, this may be understood as what can
appear as the body of a Prolog clause or what can be written as a user
query. From a logical perspective, the syntax for goals in Prolog is
the following: they can be atomic formulas or conjunctions or
disjunctions of simpler goals. In the context of Prolog, conjunctions
are written using commas and disjunctions are written using
semicolons.  Although no explicit syntax is provided for this purpose,
existential quantification may also be present in goals. Thus, a
clause of the form $\allx{x} (B(x) \supset H)$, written as {\tt H :-
B(X)} in Prolog, is equivalent (in classical logic) to $(\somex{x}
B(x) \supset H)$ if $x$ does not appear in $H$.

The language whose implementation we wish to consider is one in which
this set of logical symbols is extended to include implications and
universal quantifiers. In this language, formulas such as $F \supset
G$ and $\allx{x} G$ will be permitted as goals, provided $G$ is itself
a goal. The intended semantics of these two new operations is the
following. A goal of the form $F \supset G$ is to be solved by adding
$F$ to the current program and then solving $G$. This places a
constraint on $F$: it should have the structure of a conjunction of
program clauses. Given this understanding, implications provide a
device for giving program clauses a scope.  Thus, $F$ is to be
available only in the course of solving $G$. As for a goal of the form
$\allx{x} G$, it is intended to be solved by instantiating $x$ in $G$
with a new constant $c$ and then solving the resulting goal.
Interpreted in this fashion, the universal quantifier provides a means
for limiting the availability of names. The universally quantified
variable is, in fact, a name that is visible only within the scope of
the quantifier.

Based on the informal understanding of the new symbols, it is not
difficult to imagine that their addition to Prolog might be valuable.
A problem with Prolog is that there is no structure to its program and
name spaces. A program is a monolithic piece of code and all the
predicates defined and constants used in one place are visible
everywhere else. It is well appreciated that this is an undesirable
characteristic for a programming language. Implications and universal
quantifiers provide a means for introducing some structure.

\subsection{Lexical Scoping}

An example illustrating the problem mentioned above is provided by
auxiliary definitions. Thus, consider the definition of the reverse
relation for lists in Prolog. A naive definition of this relation is
provided by the clauses\footnote{In the examples in this section, we
use standard Prolog syntax, mixed with the obvious syntax for
implications and universal quantifiers. The reader unfamiliar with
Prolog syntax is referred to a Prolog text such as \cite{CM84}.}
\begin{exmple}
\> {\tt rev([],[]).}\\
\> {\tt rev([X|L1],L2) :- rev(L1,L3), append(L3,[X],L2).}
\end{exmple}
\noindent where {\tt append} is defined by the  usual set of clauses
for appending lists. As is well known, this definition of the reverse
relation is inefficient. The execution time of the {\tt append}
program is linear in the length of the list that is its first
argument. Its repeated invocation in the course of reversing a list
results in a program that takes time that is quadratic in the length
of the input list.

A more efficient reverse program can be written by using the idea of
an accumulator. The following definition of {\tt rev} embodies this
idea:
\begin{exmple}
\> {\tt rev(L1,L2) :- rev\_aux(L1,L2,[]).}\\
\\
\> {\tt rev\_aux([],L,L).}\\
\> {\tt rev\_aux([X|L1],L2,L3) :- rev\_aux(L1,L2,[X|L3]).}
\end{exmple}
\noindent The declarative interpretation of {\tt rev\_aux} here is
that it is true of three lists if the second is the result of
appending the reverse of the first to the last. The point to note with
this example is that {\tt rev\_aux} is an extremely specialized
predicate whose only purpose for existence is its usefulness in
defining {\tt rev}. However, its definition in Prolog occurs at the
same level as that of {\tt rev}. There are at least two undesirable
consequences of this. First, a scan of the program does not suffice
for determining what role is played by each predicate defined in it.
Second, it is possible for the name {\tt rev\_aux} to be confused with
the name of some other relation that is defined at this level, thereby
producing a mixture of definitions.

Permitting implications in goals provides a means for solving some
of these problems. The definition of {\tt rev\_aux} can be made
``local'' to that of {\tt rev} as indicated below:
\begin{tabbing}
\hspace{2em}\={\tt ((}\=\hspace{5em}\=\kill
\> {\tt rev(L1,L2) :-}\\
\> {\tt ((($\all$L$\,$rev\_aux([],L,L)) $\land$}\\
\>\> {\tt
($\all$X$\,\all$L1$\,\all$L2$\,\all$L3$\,$(rev\_aux([X|L1],L2,L3) :- 
rev\_aux(L1,L2,[X|L3]))))} \\
\>\>\> $\supset$ {\tt rev\_aux(L1,L2,[])).}
\end{tabbing}
\noindent As an explanation of the syntax of this clause, it is
obtained by moving the two clauses defining {\tt rev\_aux} into the
body of the clause defining {\tt rev}.  When appearing in the body of
a clause, a set of clauses must be represented by a conjunction and
the quantification of variables in these clauses (that was earlier
left implicit) must now be made explicit.  Factoring this in should
make the structure of the clause above clear.  The following points
might be observed with regard to this modified definition of {\tt
rev}. First, the clauses defining {\tt rev\_aux} are not available at
the top-level. Thus, these clauses will not affect the meaning of this
predicate if it is defined through some other clauses at that level.
Second, an additional structure is added to the program that helps in
understanding the purpose of its parts. For example, it is clear
merely from looking at the clause above that {\tt rev\_aux} must be an
auxiliary definition for {\tt rev}. Finally, while the definition of
{\tt rev\_aux} is not available at the top-level, the semantics
described for implication ensures that it will become 
available in the course of solving the body of {\tt rev}.  Thus,
consider the invocation of the goal {\tt rev([1,2,3],L)}. This results
in the goal {\tt 
rev\_aux([1,2,3],L,[])} being invoked after the program has been
dynamically augmented with the formula
\begin{exmple}
\> {\tt ($\all$L$\,$rev\_aux([],L,L)) $\land$}\\
\> {\tt ($\all$X$\,\all$L1$\,\all$L2$\,\all$L3$\,$(rev\_aux([X|L1],L2,L3) :-
rev\_aux(L1,L2,[X|L3])))}
\end{exmple}
\noindent The universal quantifiers and conjunction can now be made
implicit, revealing that the desired definition of {\tt rev\_aux} is
indeed available at this stage.

The use of an implication in the body of a program clause thus provides
the effect of block structuring. This gives meaning to the notions of
global and local variables within logic programming.  As an example,
using a global variable, we can eliminate the ``result'' argument of
{\tt rev\_aux} in the definition of {\tt rev}, and use the following
definition instead:
\begin{tabbing}
\hspace{2em}\=\hspace{2em}\={\tt ((}\=\hspace{5em}\=\kill
\> {\tt rev(L1,L2) :-}\\
\>\> {\tt ((rev\_aux([],L2) $\land$}\\
\>\>\> {\tt ($\all$X$\,\all$L1$\,\all$L3$\,$(rev\_aux([X|L1],L3) :-
rev\_aux(L1,[X|L3]))))} \\
\>\>\>\> $\supset$ {\tt rev\_aux(L1,[])).}
\end{tabbing}
\noindent Notice that the variable {\tt L2} is ``shared'' between {\tt
rev} and {\tt rev\_aux}. As can be seen from tracing the computation
involved in a query such as {\tt rev([1,2,3],L)}, this variable
eventually provides a means for communicating the result out to the
top-level.  Communication in the other direction --- a standard fare
in a functional programming language such as ML --- can also occur and
has its uses.  In either case, we note that the execution of a query
may require the addition of a special kind of clause, in particular, a
clause with ``tied'' variables, to the program. For example, the query 
{\tt rev([1,2,3],L)} would result in the clauses
\begin{exmple}
\> {\tt rev\_aux([],L2).}\\
\> {\tt rev\_aux([X|L1],L3) :- rev\_aux(L1,[X|L3]).}
\end{exmple}
\noindent being added to the program. Following the earlier
suggestion, we have dropped the quantifiers and the conjunction.
Note, however, that the variable {\tt L2} that appears in the first
clause is {\it not} universally quantified over the clause. Rather, it
has a binding determined dynamically at the point that it is added to
the program and is in fact identical to the variable {\tt
L} in the query.

While the use of an implication goal helps solve some of the problems
mentioned in connection with the initial definition of {\tt rev}, 
one problem still remains. The meaning of the
predicate {\tt rev\_aux} inside the body of {\tt rev} is not insulated
from definitions in existence outside the body.  Thus, if the global
program contains other clauses defining {\tt rev\_aux}, the invocation
of the implication goal does not cause a replacement of these by two
new clauses but, rather, only an addition of the two clauses to the
existing collection. This might be the desired effect in certain
situations but clearly not in the present one.

The problem under consideration can be viewed as one of limiting the
scope of the name {\tt rev\_aux} and can be solved as such
by using a universal quantifier. In particular, the
definition of {\tt rev} can be rewritten as follows:
\begin{tabbing}
\hspace{2em}\={\tt ($\all$rev\_aux$\,$((}\=\hspace{2em}\=\kill
\> {\tt rev(L1,L2) :-}\\
\> {\tt ($\all$rev\_aux$\,$((rev\_aux([],L2) $\land$}\\
\>\> {\tt ($\all$X$\,\all$L1$\,\all$L3$\,$(rev\_aux([X|L1],L3) :-
rev\_aux(L1,[X|L3]))))} \\
\>\>\> $\supset$ {\tt rev\_aux(L1,[]))).}
\end{tabbing}
\noindent To understand this definition, let us suppose that the 
goal {\tt rev([1,2,3],L)} is invoked. This leads to an 
attempt to solve the goal
\begin{tabbing}
\hspace{2em}\={\tt $\all$rev\_aux$\,$((}\=\hspace{2em}\=\kill
\> {\tt $\all$rev\_aux$\,$((rev\_aux([],L2) $\land$}\\
\>\> {\tt ($\all$X$\,\all$L1$\,\all$L3$\,$(rev\_aux([X|L1],L3) :-
rev\_aux(L1,[X|L3]))))} \\
\>\>\> $\supset$ {\tt rev\_aux(L1,[])).}
\end{tabbing}
\noindent The indicated semantics of the universal quantifier dictates
picking a new name for {\tt rev\_aux} and then solving the
instantiation of the given query with this name. Once a name is
picked, the remainder of the computation proceeds as before. However,
the fact that a new name is chosen for {\tt rev\_aux} ensures the
desired insulation from the effects of outside definitions.

\subsection{Data Abstraction}

The universal quantifier is used in the above example to hide the name
of a predicate. In a similar fashion, it may be used to hide the names
of function and constant symbols. These symbols serve to determine the
representation of data in logic programming. The ability to hide their
names therefore has the potential of supporting data abstraction. 

To illustrate this possibility, let us assume that we wish to develop
a program that uses a store. A program of this sort may be one that
carries out a graph search. Now, the development of this program can
be divided into two conceptually different tasks: (a)~the
implementation of graph search using an abstract model of the store
and (b)~the implementation of the store.  From the perspective of the
first task, we may look upon a store as being given by three
operations: $empty(S)$ that initializes $S$ to the empty store,
$remove(X,S_1,S_2)$ that produces the store $S_2$ by removing the item
$X$ from $S_1$ and $add(X,S_1,S_2)$ that produces the store $S_2$ by
adding item $X$ to $S_1$.  An implementation of graph search can now
be provided that makes no assumptions concerning the actual
implementation of these operations.

While this kind of data abstraction might be used at a conceptual
level, no language-level support is provided for it in Prolog. For
example, let us suppose that the store is represented by a stack
and the operations mentioned above are implemented through the
following clauses: 
\begin{exmple}
\> {\tt empty(emp).}\\
\> {\tt remove(X,stk(X,S),S).}\\
\> {\tt add(X,S,stk(X,S)).}
\end{exmple}
\noindent Despite the programmer's best intentions, the actual
representation of the stack, embodied in the symbols {\tt emp} and
{\tt stk}, is visible everywhere in the program and may be freely used
in the procedures that implement graph search. We also observe that
the predicates implementing the operations on the store are visible at
the top-level instead of being available only within the graph search
procedures.

Universal quantifiers and implications can be used to alleviate both
the problems mentioned above. Thus, let us assume that the store is
in fact needed for implementing graph search and that interface to
the graph search procedures is provided through a predicate of one
argument called {\tt graph\_search}. Then the definition
of the store may be relativized to the invocation of {\tt
graph\_search} by using the following query:
\begin{tabbing}
\hspace{2em}\={\tt $\all$emp$\,\all$stk$\,$((}\=\hspace{2em}\=\kill
\> {\tt $\all$emp$\,\all$stk$\,$((empty(emp) $\land$ }\\
\>\> {\tt ($\all$X$\,\all$S$\,$remove(X,stk(X,S),S)) $\land$}\\
\>\> {\tt ($\all$X$\,\all$S$\,$add(X,S,stk(X,S))))}\\
\>\>\> {\tt $\supset$ graph\_search(Solution))}
\end{tabbing}
\noindent Solving this query requires introducing new names for
the quantified variables {\tt emp} and {\tt stk}, thereby ensuring
that the ``names'' {\tt emp} and {\tt stk} that are used within the
implementation of the store are not confused with names appearing
anywhere else in the program. The semantics of implication ensures
that the operations on the store are defined at the time the procedure
{\tt graph\_search} is invoked, and hence they may be freely used
within this procedure and its auxiliary procedures.  Notice that these
procedures cannot inspect the representation of the store; in
particular, they cannot access (the new constants that replace) {\tt
emp} and {\tt stk} directly. However, they can still use the store and
can communicate through ``store valued'' variables. Note also that
there is a sense of modularity to the code presented.  The procedures
implementing the store operations can be replaced by a different
implementation without affecting the usability of the graph search
procedures. 

The various ideas described here show that implications and universal
quantifiers can be used to realize notions of modules and abstract
datatypes in logic 
programming. A fuller development of these ideas can be found in
\cite{Mil89} and \cite{Mil89jlp}.

\subsection{Metalanguage Aspects}

Prolog has certain features that make it a natural choice for
prototyping reasoning systems: it supports the idea of search in an
intrinsic way and its embodiment of first-order terms and unification
leads to convenient ways for representing and manipulating the objects
that are to be reasoned about. However, there are ways in which its
abilities in this direction can be improved. For instance, it has been
argued (\eg, see \cite{MN87slp,PE88pldi}) that using lambda terms
instead of first-order terms provides for an even better
representation of the objects that are to be manipulated. More
relevant to the present paper is the addition of the search primitives
contained in the new logical symbols being considered. One scenario
that occurs frequently in reasoning tasks is that of making an
assumption and then trying to reach a conclusion. This kind of {\it
hypothetical} reasoning is supported very naturally by implication,
given our interpretation of this symbol. Another paradigm that
is useful is that of introducing a new object and then determining if
a given statement is true of it. This is the basis, for instance, of
universal generalization. Universal quantifiers in goals provide a
means for realizing this paradigm.

We illustrate the above observations by considering the task of type
inference for lambda terms.\footnote{We assume a familiarity in the
rest of this section with basic lambda calculus notions. The reader
unfamiliar with these may consult \cite{HS86} or some similar
source.}\ These terms are constructed from constants and variables
using the operations of abstraction and application. We assume that
the types of constants are previously specified. The types of
variables are determined by an {\it environment}. An arbitrary lambda
term can then be inferred to be of a certain type relative to an
environment $\Gamma$ by using the following rules: 

\begin{romanlist}

\item An occurrence of a constant has as a type any instance of
the type specified for the constant.

\item Every occurrence of a variable has as its (sole) type the one
assigned to the variable by $\Gamma$.

\item If $t_1$ and $t_2$ have $\alpha \ra \beta$ and $\alpha$ as types
relative to $\Gamma$, then $(t_1 \app t_2)$ has $\beta$ as a type
relative to $\Gamma$.

\item If $t$ has $\beta$ as a type relative to an environment that is
like $\Gamma$ except that that it assigns the type $\alpha$ to $x$,
then $\lambdax{x}t$ has $\alpha \ra \beta$ as a type relative to
$\Gamma$. 

\end{romanlist}

\noindent Types are assumed to be polymorphic here, and are
represented by first-order expressions with the single binary infix
function symbol $\ra$ and a collection of constant symbols that
represent the primitive types.

Suppose now that we wish to write a logic program that infers types
for closed lambda terms. This program will need, first of all, to
associate types with constants. These associations can be represented
through facts or atomic clauses. The program will also need to
represent an environment. Since our interest is in inferring types for
closed terms, it is necessary only to maintain the types assigned to
bound variables by the environment. Thus, the environment may also be
represented by means of a set of facts, with implication goals being
used to add to this set at the point where abstractions are
encountered. To provide concreteness to this discussion, let us
suppose that the only constants available are $1$ and $+$ of type
$int$ and $int \ra (int \ra int)$ respectively. Then the following
program represents an attempt at implementing type inference using
these ideas:
\begin{exmple}
\> {\tt type\_of(1,int).}\\
\> {\tt type\_of(+, int $\ra$ (int $\ra$ int)).}\\
\> {\tt type\_of(app(E1,E2),T1) :- type\_of(E1,T2 $\ra$ T1),
type\_of(E2,T2).}\\ 
\> {\tt type\_of(abst(X,E),T1 $\ra$ T2) :- (type\_of(X,T1) $\supset$
type\_of(E,T2)).} 
\end{exmple}
\noindent We have assumed a first-order representation of lambda terms
in this program, with abstraction and application being represented by
the binary function symbols {\tt abst} and {\tt app} respectively, and
(object-language) constants by suitably chosen constant symbols.

A question that is not yet settled in connection with the above
program is whether variables in lambda
terms are to be represented by variables or constants of the
programming language.  A brief consideration of this question leads to
the conclusion that (metalanguage) variables are not the right choice:
using such a representation would permit, for instance, the erroneous
inference that $\lambdax{x}\lambdax{y}((+ \app x) \app y)$ has $\alpha
\ra (\beta \ra int)$ as one of its types for any choice of types for
$\alpha$ and $\beta$. 
Unfortunately, there is a problem with the program shown even if 
constants are used to represent the variables in lambda terms. The
source of this problem is that the same 
variable name may be used for different abstractions in a given lambda 
term and, in this case, an inner abstraction is intended to ``hide''
the outer abstraction. It is by virtue 
of this convention that a term such as $\lambdax{v} \lambdax{v}( (+
\app v) \app (v \app v))$ is deemed to be ill-typed. However, this
hiding effect is not realized by our program. Thus, assuming that
lambda term variables are represented by constants of the
same name, the term $\lambdax{v} \lambdax{v}( (+ \app v) \app (v \app
v))$ will be judged by our program to have 
$(int \ra int) \ra (int \ra int)$ as one of its types. 

The problem can be solved by using universal quantifiers. However, we
need to change our representation of lambda terms before we can
describe this solution. To begin with, we assume that the data
structures of our language are themselves provided by lambda terms and
not first-order terms. We do this because we need an encoding of
substitution in the solution we provide, and using lambda terms as
data structures leads to this being available as a primitive
operation. Now, we represent an object-language expression such as
$\lambdax{x} E$ by {\tt abst($\lambda$x E)} where {\tt E} is the
translation of $E$ (with $x$ replaced by {\tt x}).  The scheme for
representing applications remains unchanged.  Using this
representation of lambda terms, a correct type inference program is
given by the following clauses:
\begin{exmple}
\> {\tt type\_of(1,int).}\\
\> {\tt type\_of(+, int $\ra$ (int $\ra$ int)).}\\
\> {\tt type\_of(app(E1,E2),T1) :-}\\
\>\>{\tt type\_of(E1,T2 $\ra$ T1), type\_of(E2,T2).}\\ 
\> {\tt type\_of(abst(E),T1 $\ra$ T2) :- }\\
\>\>{\tt ($\all$x$\,$(type\_of(x,T1) $\supset$ type\_of(E(x),T2))).} 
\end{exmple}
\noindent The manner in which the universal quantifier in the body of
the third clause serves to introduce a new constant dynamically should
be noted in this example. This constant must be substituted into the
body of the abstraction. By virtue of our representation of terms,
this effect is produced by the application of {\tt E} to the
quantified variable {\tt x}. This application is written in the
program above as {\tt E(x)}.

The examples considered in this section are simple ones, intended only
to bring out the semantics of the new logical symbols and the value of
their inclusion in logic programming. More extensive examples may be
found in various places in the literature. (See, for example,
\cite{Felty89phd,Hannan90,Mil89,Mil89jlp,PM90iclp}.) 
In the following sections we provide a precise definition of a logical
language that includes implications and universal quantifiers in goals
and we examine the implementation of this language. The language that
we consider is a first-order one and does not explicitly cover all the
examples presented here. This simplification is chosen largely for
expository reasons and nothing essential to the implementation of
implications and universal quantifiers in goals is left out by it.
Towards bringing this point out, we indicate in Section
\ref{sec:higherorder} the additional devices necessary for handling
the higher-order aspects present in the examples of this section.

\section{AN EXTENDED LANGUAGE AND THE PROBLEMS IN ITS
IMPLEMENTATION}\label{sec:fohh}  

A language that utilizes implications and universal quantifiers can be
described as an extension to the language based on Horn clauses.  In
describing this extension, we need to adopt a somewhat more general
view of Horn clauses than is usual.  Based on the methodology
developed in \cite{MNPS91}, the logic underlying a logic programming
language may be characterized by two classes of formulas: the $G$
formulas that function as goals or queries, and the $D$ formulas that
fill the role of program or, to use a terminology common in
discussions of Horn clauses, definite clauses. In this context, the
programming framework provided by Horn clauses is defined by the $G$
and $D$ formulas given by the following syntax rules in which $A$
represents an atomic formula:
\[
\begin{tabular}{l}
$G ::= A \sep (G\land G) \sep (G\lor G) \sep (\somex{x}G)$,\\
$D ::= A \sep (G \supset A) \sep (\allx{x} D)$.\\
\end{tabular} \]
The parentheses that surround expressions in these and other syntax
rules are included to ensure unique readability and may be omitted if
doing so does not cause an ambiguity. Now, the formulas described
above are related to Horn clauses in the following sense:
within the setting of classical logic, the negation of a $G$ formula
is equivalent to a set of negative Horn clauses and, similarly, a $D$
formula is equivalent to a set of positive Horn clauses. The syntax
adopted here is motivated by its greater proximity to actual
programming realizations, its amenability to extensions and our use of
derivability, as opposed to refutability, as the primitive semantic
notion.

In the framework of \cite{MNPS91}, the task of programming consists of
describing a set of relationships between objects through a collection
of closed program clauses thought of as a program, and of querying
such a specification through goals.  From a logical perspective, this
viewpoint is justified only if the task of answering a query can be
equated with the notion of constructing a proof for the query from the
given program. In the context of Horn clauses, use can be made of
either classical or intuitionistic provability to satisfy this
requirement. Both derivability relations validate the following recipe
for solving a closed goal $G$ given a program $\Pscr$:

\begin{numberedlist}
\item if $G$ is $G_1 \land G_2$ then try to solve it by solving both 
$G_1$ and $G_2$, 

\item if $G$ is $G_1 \lor G_2$, then try to solve it by solving either 
$G_1$ or $G_2$, 

\item if $G$ is $\somex{x}G_1$, then try to solve it by solving
$\subs{t}{x}{G_1}$ for some closed term $t$, and 

\item if $G$ is an atom, then try to solve it (a)~by determining that
it is an instance of a program clause in $\Pscr$, or (b)~by finding an
instance $G_1 \supset G$ of a program clause in $\Pscr$ and trying to 
solve $G_1$. 
\end{numberedlist}

\noindent The program is assumed to be fixed throughout the above
description, and the notation $\subs{t}{x}{G}$ is used to denote the
result of replacing every free occurrence of $x$ in $G$ by $t$, taking
care, of course, to avoid the inadvertent capture of free variables.
The most interesting aspect of the above recipe is that it permits the
connectives and quantifiers in goals to be interpreted dually as
search primitives. Under this interpretation, $\lor$ and $\land$
respectively specify OR and AND branches in a search and the
existential quantifier specifies an infinite OR branch with the
branches parameterized by closed terms.  The behavior of existential
quantifiers also permits ``answers'' to be extracted from
computations: a goal with free variables may be interpreted as a
request to solve the existential closure of the formula and to produce
instantiations for the introduced quantifiers that lead to successful
solutions.

The extended language that we desire must permit implications and
universal quantifiers in goals. These additions are incorporated into 
a language that is based on first-order hereditary Harrop or {\em
fohh} formulas \cite{MNPS91}. The syntax of goals and program clauses
in such a language is given by the $G$ and $D$ formulas
described by the following rules:
\[
\begin{tabular}{l}
$G ::= A \sep (G\land G) \sep (G\lor G) \sep (\somex{x}G) \sep (\dset
\supset G) \sep (\allx{x} G)$,\\
$\dset ::= D \sep (D \land \dset)$,\ {\rm and} \\
$D ::= A \sep (G \supset A) \sep (\allx{x}D)$.\\
\end{tabular} \]
Note that the implications that are permitted in goals in this
extended language are limited --- conjunctions of $D$ formulas must 
appear on the left and $G$ formulas on the right. However, this
restriction is in keeping with our informal discussion in the previous
section. 

Our desire is to interpret implication and universal quantification as 
scoping mechanisms with respect to program clauses and names
respectively. This is exactly the effect we obtain if the idea of
solving a goal with respect to a program is clarified using the notion 
of intuitionistic provability. In particular, if $\Pscr$ is the 
program, then the following additions need to be made to the earlier
recipe to get one for solving a closed goal $G$ in the new language:

\begin{numberedlist}
\item[\hss(5)\enspace]if $G$ is $(D_1 \land \ldots \land D_n) \supset
G_1$, then try to solve it by solving $G_1$ using ${\Pscr} \cup
\{D_1,\ldots,D_n\}$ as the program instead of $\cal P$, and 

\item[\hss(6)\enspace]if $G$ is $\allx{x} G_1$, then try to solve it
by solving $\subs{c}{x}{G_1}$ for some new constant $c$. 
\end{numberedlist}

The recipes for solving a goal from a program that are provided above
are useful in understanding the nature of computation in the languages
that are based on Horn clauses and on {\em fohh} formulas.  They also
provide some indication of how computations might actually be carried
out. However, they are not complete from this perspective.  One
problem is that the instruction for solving existential goals assumes
an oracle for picking the ``right'' instantiation for the quantifier. 
Similarly, choices have to be made concerning the disjunct to be
solved in a disjunctive goal and the program clause to be used in
solving an atomic goal. In each of these cases, some machinery is
needed in addition to the basic instruction to support the making of
these choices.

\ignore{ The techniques that are used in conjunction with our recipe
in implementing the language based on Horn clauses are, by now, quite
standard.}
The additional machinery that suffices for implementing the Horn
clause language is, by now, quite standard.
The problem with existential quantifiers is dealt with by delaying the
actual instantiations of such quantifiers till such time that
information is available for making an appropriate choice.  This
effect is achieved by replacing the quantified variables by
placeholders whose values are determined later through the process of
unification.  Thus, a goal such as $\somex{x}G(x)$ is transformed into
one of the form $G(X)$ where $X$ is a new logic variable that may be
instantiated at a later stage.  In attempting to solve an atomic goal
$A$, we look for a definite clause $\allx{{\bar y}} (G' \supset A')$
such that $A$ unifies with the atomic formula that results from $A'$
by replacing the universally quantified variables with new logic
variables.  If such a clause is found, the next task becomes that of
solving the resulting instance of $G'$.  The approach that is used to
deal with the other forms of nondeterminism is to assume an implicit
ordering of choices and to implement a depth-first search with the
possibility of backtracking; thus, disjunctive goals are considered in
left-to-right order and program clauses are used in the order of
presentation. A final point to note is that much of the unification,
the processing of the search primitives, and the sequencing through
program clauses can be compiled within this framework. These various
observations are in fact used in WAM-based approaches to provide
extremely efficient implementations for the programming paradigm based 
on Horn clauses.

Our desire in this paper is to extend these methods to obtain a
satisfactory implementation of a programming language based on {\em
fohh} formulas. It may appear that such an extension can be easily
obtained: in order to deal with universal quantifiers, we merely need
to consider the instantiation of a goal with a newly generated
constant and to deal with implications we only need a mechanism for
adding program clauses to a program.  However, an implementation based 
solely on this view would be both incorrect and inadequate.

The suggestion for dealing with universal quantifiers ignores
interactions with the scheme being built upon and would be erroneous
if executed naively. The source of the problem is that universal and
existential quantifiers can appear in arbitrary orders in the goals
that are of interest. For example, consider the task of solving the
goal $\somex{x} \allx{y} p(x,y)$, where $p$ is a predicate symbol.
Using the scheme suggested, we may reduce this task to that of solving
the ``goal'' $p(X,c)$ where $c$ is a new constant and $X$ is a logic
variable.  However, unification cannot be used in an unqualified
fashion in solving the new goal because any instantiation that is
determined for $X$ must not contain $c$ in it.  Thus, suppose that we
attempt to solve the goal $\somex{x}\allx{y}p(x,y)$ in the context of
a program containing the clause $\allx{x}p(x,x)$. If care is not
exercised, an incorrect derivation for the goal may be constructed by
(indirectly) instantiating $X$ to $c$.\footnote{In the context of
classical logic, universal quantifiers can be eliminated by using
Skolem functions of the existentially quantified variables within
whose quantifier scope they appear. (Note that the roles of
quantifiers are reversed in the refutability setting.)  Incorrect
instantiations of the kind discussed above will then be blocked by the
process of ``occurs-checking'' in unification.  Unfortunately, as
discussed in \cite{Nad92int}, the problem cannot be dealt with in the
present context in a similar ``static'' fashion.  Some feeling for
this might be obtained by trying to determine how the static process
ought to work in conjunction with the goal $(\allx{x} p(x) \supset q)
\supset \somex{x}(p(x) \supset q)$, noting that this goal should not
succeed. However, a dynamic form of Skolemization does work even in
this context. The solution used in this paper captures the constraints
dynamic Skolemization is designed to capture in a much more direct
fashion.}

In providing a satisfactory treatment of implications in goals, there
are several aspects that require a detailed consideration. First, we
observe that in its presence the program being used cannot be left
implicit. Thus, consider solving the goal $(D_1 \supset G_1)
\land (D_2 \supset G_2)$ from a program $\Pscr$. This task eventually
requires two different programs, \ie, $\Pscr \cup \{D_1\}$ and $\Pscr
\cup \{D_2\}$, to be used in solving the goals $G_1$ and $G_2$.
An acceptable implementation should not require the explicit
construction of two separate programs but rather should support the
realization of the two different contexts through a process of gradual
addition and removal of code. Such a scheme can actually be supported
and the implementation we describe later even permits the compilation
of program clauses that are to be added to the original program.
However, the interaction of backtracking with this approach requires
bookkeeping devices of some sophistication. To see why this is the
case, consider solving the goal $\somex{x} ((D \supset G_1(x)) \land
G_2(x))$ from the program $\Pscr$. Under the scheme being
considered, we would first have to augment the program with $D$ and
attempt to solve the goal $G_1(X)$, where $X$ is the logic variable
introduced for the existential quantifier. A successful solution would
determine a binding for $X$. An attempt would now have to be made to
solve the appropriate instance of $G_2(X)$ after removing $D$ from the 
program.  Assume now that, with the instantiation found for $X$,
$G_2(X)$ cannot be solved. The requirement then is to look for another
solution to $G_1(X)$. However, such a solution must be sought in the
context of the relevant program; in particular, the program clause $D$
must be reinstated and any additions made in the course of trying to
solve $G_2(X)$ must be removed. In general, we see that backtracking
may require the program that is to be used to be changed
substantially, and mechanisms have to be provided for realizing such
switches in context in an efficient manner.

The final problem concerns the presence of tied variables in program
clauses. One situation in which this arises is that when existential
quantifiers are used in conjunction with implications.  Thus, consider
solving the goal $\somex{x} (p(x) \supset g(x))$ where $p$ and $g$ are
predicate names. Assuming that $x$ is replaced by the logic variable
$X$ and implication is dealt with in the manner required, we would
have to solve the goal $g(X)$ with respect to a program that contains
the clause $p(X)$. Notice that the variable that occurs in $p(X)$ is
different from the variables that usually occur in program clauses: it
cannot be instantiated in arbitrary fashion but rather only in one
particular way that is also consistent with the instantiation for the
occurrence of the same variable in the goal $g(X)$. To appreciate this
aspect completely, consider solving the given goal from a program
containing the clauses $q(a)$ and $\allx{x} ((q(x) \land p(b)) \supset
g(x))$, assuming that $a$ and $b$ are constants and $q$ is a predicate
name. It may appear at first that the goal should succeed in this
context. Thus, we may backchain on the second clause in the original
program to solve $g(X)$, producing the subgoals $q(X)$ and $p(b)$.
The subgoal $q(X)$ might be solved by using the clause $q(a)$, and the
subgoal $p(b)$ may apparently be solved by using the program clause
$p(X)$. Such a solution would in reality be erroneous: the variable in
the program clause $p(X)$ is tied to the one in the goal $g(X)$ and,
thus, this ``solution'' involves instantiating the logic variable $X$
simultaneously with $a$ and $b$. More generally, we see that a
suitable implementation of our language must contain mechanisms for
distinguishing between variables of two different kinds that might now
appear in programs and also for dealing with the new kind of
variables.

In the remainder of this paper we develop methods for dealing with the
various new implementation problems that arise in the context of a
language that is based on {\em fohh} formulas. We shall describe these
methods as extensions to the 
machinery already present in the WAM.  Two questions need to be
answered in justifying this approach: why is the WAM used as a
starting point and might not a metaprogramming approach, perhaps using
the impure predicates present in Prolog, yield a satisfactory result
as well? We have discussed in this section the manner in which a
language that is based on {\em fohh} formulas builds on one that is
based on Horn clauses and have
also motivated the use of several mechanisms present in
implementations of the latter in obtaining an implementation of the
former. This discussion provides a strong argument for utilizing the
structure of the WAM in implementing the language that is presently of
interest.  Concerning the second question, we observe first that there
are substantial new issues that need to be considered prior to an
implementation and part of the objective of the ensuing sections is to
study these issues and to suggest mechanisms for dealing with them.
These discussions are thus relevant even if a metaprogramming approach
is to be used. We further note that certain situations arise in the
processing of our language that are alien to the setting of Horn
clauses. 
These include the introduction of new constants through universal
quantifiers and the possibility of sharing variables between clauses
and goals. A metaprogramming approach does not offer any natural
advantages in dealing with these situations and, therefore, we feel
that the specific approach that is adopted here is justified.

\section{AN ABSTRACT INTERPRETER}\label{sec:interp}

We deal first with the problem arising from existential and universal
quantifiers appearing in mixed order in goals. We describe in this
section an abstract interpreter for our language that incorporates a
solution to this problem within it. The source of the problem is that
the set of terms available at the point where a logic variable is
introduced may be different from that in existence at a later stage in
the computation and that the substitutions that are made for the
variable must be restricted to the former set. For example, let us
suppose that our language has one unary function symbol $f$ and one
constant symbol $a$ and then consider the attempt to solve the goal
$\somex{x}\allx{y} p(x,y)$. Using the steps outlined in the previous
section, this results in an attempt to solve the goal $p(X,c)$ where
$c$ is a new constant. Notice that at this stage our universe of terms
has been expanded by the addition of the constant symbol $c$. However,
it is the old collection of terms, that obtained by using $f$ and $a$
and variables, that determines acceptable substitutions for $X$.

A naive approach to ensuring that only legitimate instantiations are
considered for logic variables involves tagging each of these
variables with the set of constants that are permitted to appear in
terms instantiating it. This set can then be used in an
``occurs-check'' during unification in order to determine the
acceptability of proposed substitutions.  Fortunately, the different
sets of constant symbols constitute a hierarchy of universes and a
practical realization of this idea can be obtained by using a
numerical tag with each constant and logic variable. The level $1$
universe consists of all the constant symbols that appear in the
program clauses and the original goal. These symbols may be tagged by
$1$ to indicate their position in the hierarchy. Each time a universal
quantifier is encountered, a new constant must be introduced, giving
rise to the next universe in the hierarchy. This requirement can be
accounted for by increasing the ``universe index'' by $1$ and
introducing a new constant tagged with this index. The collection of
constants at the new level thus consists of all those constants tagged
with a number less than or equal to that level. When an existential
quantifier is encountered, it is instantiated by a logic variable.
This variable may be tagged with the current value of the universe
index, the intended interpretation of the tag being that a term may be
substituted for the variable only if all the constants appearing in
the term have a smaller or identical tag.

The actual use of the tags occurs in the course of unification
and consists of the following. The process of unification
culminates with an attempt to instantiate a logic variable with a
term. In the present context, this would amount to an attempt to set a
variable $X$ with a tag $i$ to a term $t$. Before such an
instantiation is permitted, a consistency check must be performed on
tags in addition to the usual occurs-check: it must be determined that
$t$ does not contain any constants with a tag value greater than $i$.
Actually, one additional device must be incorporated into this basic
scheme to make it work correctly. Suppose that we have determined that
it is acceptable to set $X$ to $t$. Before actually doing this, it is
necessary to change to $i$ the tags on variables appearing in $t$ that
have a value greater than $i$. This is required in order to prevent a
later instantiation of these variables from violating the restrictions
on instantiations for $ X$.  As an illustration, suppose that our
program consists of the single clause $\allx{z} (q(z) \supset
p(d(z)))$ and that we are trying to solve the goal $\somex{x}\allx{y}
(q(y) \supset p(x))$. After the quantifiers are processed, the goal
becomes $(q(c^{2}) \supset p(X^{1})$; we assume that numerical tags
are associated with constants and logic variables in the manner just
described and we depict these tags as superscripts on the relevant
symbols. The attempt to solve this goal results, in turn, in an
attempt to solve $p(X^{1})$ from a program containing the clauses
$q(c^{2})$ and $\allx{z} (q(z) \supset p(d(z)))$. Backchaining on the
second clause now yields the goal $q(Z^{1})$, which fails. The
important point to note here is that failure is dependent on the tag
value of $X$ being communicated to the new logic variable $Z$ that is
used to instantiate the quantifier in the clause in question.
\ignore{ As another example, consider the goal
$\somex{x}\allx{y}\somex{z} (p(x, f(z)) \land p(y,z))$ in the context
of the program $\allx{x} p(x,x)$. Although the ``subgoals'' $\somex{x}
\allx{y} \somex{z} p(x,f(z))$ and $\somex{x} \allx{y} \somex{z}
p(y,z)$ of the given goal have solutions, the combination does not
have one.  Once again, the process of ``tag propagation'' is essential
for preventing an erroneous solution from being found for the
composite goal.}

The above discussion outlines a notion of labelled or tagged
unification that is relevant to the implementation of the language
that is based on {\em fohh} formulas.  A formal presentation of this
notion and a study of some of its properties may be found in
\cite{Nad92int}.  For our  present purposes it suffices to note that
this form of unification can be explained in a fashion similar to
first-order unification, that the notion of most general unifiers
makes sense in this context as well and that such unifiers can be
found by a process identical to that in the usual first-order case
except for the checking of tag constraints and the propagation of tags
described above. In the sequel, we relativize all the terminology
pertaining to unification to this notion in the extended sense just
described. 

The use of the ideas described above in dealing with a mixture of
quantifiers in goals calls for a method for associating tags with the
constants and logic variables appearing in such goals. We have already
described the way in which this tag is determined if the constant or
logic variable is introduced as a result of processing a quantifier.
However, processing may start with a goal that already has constants
and free variables (that eventually become logic variables) in it. In
this case a {\it tagged version} of the goal is produced by
associating the tag $1$ with these constants and variables. Similarly,
it may be necessary at some point in the computation to create an
instance of a program clause and the constants and free variables
appearing in such an instance must be tagged. An instance of this kind
will be needed when the universe index is at some value $I$ and it
constitutes a {\it new tagged instance} of the clause
relative to $I$ that is obtained

\begin{romanlist}

\item by associating the tag $1$ with each untagged constant appearing
in the clause if the clause is of the form $A$ or $G \supset A$, and 

\item by picking a new variable $w$, associating the tag $I$ with $w$
and obtaining a new tagged instance of $\subs{w}{x}D$ relative to
$I$ if the clause is of the form $\allx{x} D$. 

\end{romanlist}

\noindent We assume here that the free (alternatively, logic)
variables that appear in a program clause are already tagged. This
property holds trivially for all the clauses in the original program
since these  
are assumed to be closed and can be seen to hold for all the clauses
that arise in the course of the processing that is described below.

We now present the promised abstract interpreter.  We note that, in
this presentation, the free variables and constants in ``goals'', the
free variables in ``program clauses'', and the variables and constants
in ``substitutions'' will all be tagged. We continue to refer to these
objects as goals, program clauses and substitutions, despite this
change. Now, the possibility for implications to be present in goals
makes it necessary to consider explicitly the program clauses that are
available when a particular goal is being solved. Similarly, the
inclusion of universal quantifiers requires the solution of a goal to
be parameterized by a universe index. Thus, our abstract interpreter
will deal with tuples of the form $\langle G,\Pscr, I \rangle$, where
$G$ is a goal, $\Pscr$ is a program and $I$ is a natural number. We
shall refer to a multiset of such tuples as a decorated goal set. Let
$\Gscr$ be a decorated goal set and let $\theta$ be a substitution.
Then the abstract interpreter may transform the pair $\langle \Gscr,
\theta \rangle$ according to the following rules:

\begin{numberedlist}

\item If $\Gscr$ is ${\Gscr'} \cup \{\langle G_1 \land
G_2,\Pscr,I\rangle \}$, then by obtaining $\langle \Gscr' \cup 
\{\langle G_1,\Pscr,I \rangle, \langle G_2,\Pscr,I \rangle\},
\emptyset\rangle$.

\item If $\Gscr$ is $\Gscr' \cup \{\langle G_1 \lor
G_2,\Pscr,I\rangle \}$, then by obtaining $\langle \Gscr' \cup
\{\langle G_i,\Pscr,I \rangle\}, \emptyset\rangle$ for $i = 1$ or 
$i = 2$.

\item If $\Gscr$ is $\Gscr' \cup \{\langle \somex{x}G ,\Pscr,I\rangle 
\}$, then by obtaining $\langle \Gscr' \cup \{\langle
\subs{w}{x}G,\Pscr,I \rangle\}, \emptyset\rangle$, where $w$ is a new
variable whose associated tag is $I$.

\item In the case where $\Gscr$ is $\Gscr' \cup \{\langle (D_1 \land
  \ldots \land D_n) 
\supset G ,\Pscr,I\rangle \}$, by obtaining $\langle \Gscr' \cup 
\{\langle G,\Pscr \cup \{D_1,\ldots,D_n\},I \rangle\}, \emptyset\rangle$.

\item If $\Gscr$ is $\Gscr' \cup \{\langle \allx{x} G ,\Pscr,I\rangle
\}$, then by obtaining $\langle \Gscr' \cup 
\{\langle \subs{c}{x}G,\Pscr,I + 1 \rangle\}, \emptyset\rangle$,
where $c$ is a new constant whose associated tag is $I+1$.

\item If $\Gscr$ is $\Gscr' \cup \{\langle A ,\Pscr,I\rangle \}$ and
$G \supset A'$ is a new tagged instance relative to $I$ of a clause in
$\Pscr$ such that $A$ and $A'$ have the most general unifier $\sigma$,
then by obtaining $\langle \sigma(\Gscr' \cup \{\langle G, \Pscr,
I\rangle\}), \sigma\rangle$.\footnote{Applying a 
substitution to a set is to be interpreted as applying it to each
element and applying it to a tuple corresponds to applying it to each
formula that appears in it. Note also that the presence of quantifiers
in formulas may require renamings to carried out in the course of
applying a substitution. In an actual implementation the
representation of free variables eliminates the usual capture problems
and thus obviates renaming.}

\item If $\Gscr$ is $\Gscr' \cup \{\langle A ,\Pscr, I\rangle \}$
and $\sigma$ is a most general unifier of $A$ and a new tagged
instance of a clause in $\Pscr$ relative to $I$, then by
obtaining $\langle \sigma(\Gscr'), \sigma\rangle$. 
\end{numberedlist}

\noindent The symbol $\cup$ used in these transition rules 
denotes multiset union. The abstract interpreter for our language now
functions as follows. In attempting to solve a goal $G$ given a
program $\Pscr$, it will start off with the tuple $\langle \{\langle
G', \Pscr, 1\rangle\}, \emptyset\rangle$, where $G'$ is a tagged
version of $G$, and will transform this tuple by repeated applications
of the rules above. It will succeed if it eventually manages to obtain
a tuple of the form $\langle \emptyset,
\theta\rangle$. In this case, the sequence of tuples $\langle 
\Gscr_i, \theta_i \rangle_{1 \leq i \leq n}$ that constitutes
a successful run for the interpreter is referred to as a
{\it derivation} of $G$ from $\Pscr$, and $\theta_n \circ \ldots \circ
\theta_1$ is referred to as the associated answer substitution. 

There is an evident non-determinism in the interpreter. This
non-determinism can be factored into two forms. First, there may be a
choice concerning the tuple from the decorated goal set that is to be
processed next. Second, there may be a choice concerning the disjunct
that is to be solved if the tuple picked pertains to a disjunctive
goal and the program clause that is to be used if the tuple picked
pertains to an atomic goal. The latter kind of non-determinism is one
that we have discussed already and is manifest in the transition rules
(2), and (6) and (7) respectively.  The former kind of non-determinism
is inconsequential.  The following proposition attests to this fact
and also verifies the correctness and the adequacy of the abstract
interpreter that is described above. A proof of this proposition may
be found in
\cite{Nad92int}. 

\begin{proposition}\label{scprop}
Let $\Pscr$ be a program and let $G$ be a
goal. 

\begin{numberedlist}

\item 
If there is a derivation of $G$ from $\Pscr$ with answer
substitution $\theta$, then there is a proof in intuitionistic logic
for $\theta(G)$ from $\Pscr$.

\item 
If, for some substitution $\sigma$, there is a proof in
intuitionistic logic of $\sigma(G)$ from $\Pscr$, then there is a
derivation of $G$ from $\Pscr$ with an answer substitution $\theta$
that is more general than $\sigma$. Furthermore, such a derivation can
be obtained by picking the next tuple to be processed in an arbitrary 
fashion. 
\end{numberedlist}

\end{proposition}

The abstract interpreter has several features that makes it amenable to
a WAM-like implementation. The essential non-determinism that is
present in it is similar to that in the case of Horn clause logic and
can be handled, as usual, by a depth-first search with backtracking,
to be implemented through the use of choice point records. In contrast
to the Horn clause case, constants and variables have to be tagged and
these tags have to be utilized in determining unifiers. The universe
index will be needed in the generation of these tags and we add a
register called the {\tt UI} register to those present in the WAM for
maintaining this index. This register will be manipulated by universal
goals, being incremented on entry and decremented on successful
completion.  Backtracking may, in general, cause a switch to a context
embedded within a different number of universal quantifiers and the
register must be reset to the appropriate value in such cases.  To
facilitate such a resetting, choice point records include, in our
context, an additional field called {\tt UIP} into which the value of
the {\tt UI} register is stored at the time of creation of the record.
The use of tags in unification is, of course, quite straightforward.
From the perspective of compilation, instructions for unification need
to be modified so as to utilize the tags in the required fashion.
Although no new instructions are needed for compiling unification,
some new instructions are required for handling the effects
of universal quantifiers. The details of these aspects are discussed
in Section \ref{sec:compile}.

There is, however, one aspect of implementation that needs further
consideration.  The possible occurrence of implications in goals
requires that the solution of each goal be relativized to a program
context. In the abstract interpreter, this requirement is fulfilled by
decorating each goal with its program context. Construing such a
decoration naively will obviously not lead to an acceptable
implementation. However, it is possible to provide a stack-based
realization of changing program contexts and we discuss this issue in
the next section.

\section{DEALING WITH IMPLICATION GOALS}\label{sec:impl}

An invocation of the implication goal $D \supset G$ causes $D$ to be
added to the program before an attempt is made to solve $G$ and to be
removed from the program upon a successful completion of this attempt.
Thus, implication goals conceptually entail ``asserting'' and
``retracting'' program clauses. Invocations of such goals can be
nested inside one another and several layers of these operations may,
therefore, have to be performed during execution. However, the
assertion and retraction of program clauses follows a stacking
discipline and can, in principle, be implemented using a run-time
stack.

An actual implementation of the above conceptual model must include
devices for dealing with certain additional aspects.  One of these
aspects is the sharing of code across different versions of the
``same'' program clause that may be added to the program in the course
of solving a query. To understand what exactly is at issue, let us
suppose that our program contains the clauses $p(a)$ and $\allx{x}(((D
\supset G) \land p(x)) \supset p(f(x)))$, where $p$ is a predicate
name, $f$ is a function symbol, $D$ is a program clause and $G$ is a
goal, and then consider solving the goal $\somex{y} p(f(f(y)))$. The
goal $(D \supset G)$ will be invoked twice in the course of solving
the given goal. The clause $D$ will, therefore, have to be added to
the program twice. However, a satisfactory implementation should
maintain only one copy of the ``code'' for $D$ and use this in
realizing both additions. Adopting this approach is necessary both for
controlling the sizes of program and for supporting the compilation of
program clauses.

The sharing of code for program clauses can be easily accomplished in
the example considered above: we simply maintain one copy of the code
for $D$ and use pointers to this on the two different occasions that
it is added to the program. However, as discussed in
Section~\ref{sec:fohh}, it is possible for program clauses to contain
free variables and so this idea does not quite solve the problem in
the general case. As a specific illustration, suppose that the second
program clause in the example considered above is replaced by the
clause
\[\allx{x} (((D(x) \supset G) \land p(x)) \supset p(f(x)));\]
we assume here that $D(x)$ represents a program clause with $x$
occurring free in it. The two program clauses that are added to the
program in the course of solving the goal $\somex{y}p(f(f(y)))$ are
now $D(f(y))$ and $D(y)$. These program clauses are, in a sense,
distinct. However, they could have a considerable amount of structure
in common, and a reasonable implementation scheme should permit this
structure to be shared.

The above considerations lead naturally to a representation of a
program clause as a composite of (a pointer to) code and a set of
bindings for its free variables. Such a representation corresponds to
the idea of a closure that is used in implementations of functional
programming languages and is an enrichment to the usual WAM treatment
of program clauses. Using such a representation makes it possible to
compile both the program clauses that appear as the antecedents of
implications and the action to be taken on encountering implication
goals. In understanding how this might be done, let us assume that
programs are maintained as lists of closures that are searched
sequentially in order to determine the clauses relevant to solving
given atomic goals; this representation of programs differs from the
one used in the WAM and is also extremely naive, but we defer the
consideration of more sophisticated representations till the next
section. Now suppose that there is an occurrence in the original
program or goal of an implication goal of the form \[(D_1(\bar x_1)
\land \ldots \land D_n(\bar x_n))\supset G,\] where, for $1 \leq i
\leq n$, $D_i(\bar x_i)$ denotes a program clause and $\bar x_i$ is a
listing of the variables occurring free in it. The variables in
${\bar x}_i$ are, in fact, ones that are bound by quantifiers that
surround the implication goal in question. An invocation of this 
implication goal will, therefore, take place in a context where these 
variables have been replaced by logic variables or by generated
constants.  As we shall see in detail in Section~\ref{sec:compile},
bindings for these variables at a particular invocation can be given
by compile-time determined offsets relative to the current environment
record.  Now, the program clause given by $D_i({\bar x}_i)$ can be
compiled in the usual fashion with the exception that it should
include instructions for initializing the variables in ${\bar x}_i$ as
described below. Let $c_i$ be a pointer to this code.  The enhancement
to the program that is required at an invocation of the implication
goal can be realized by adding to it the closures $\langle
c_i,e\rangle$ for $1 \leq i \leq n$, where $e$ is a pointer to the
current environment record.  This action can itself be compiled by
statically associating with the goal a table of pointers to code for
the program clauses that constitute its antecedent.  Suppose now that
a version of the clause $D_i({\bar x}_i)$ that is given by the closure
$\langle c_i, e \rangle$ is invoked in an attempt to solve some
(sub)goal. The code that is to be executed is pointed to by $c_i$.
This code must, first of all, relativize the bindings for the
variables in $\bar x_i$ to the environment record created for the
invocation. Doing this involves copying over the bindings for these
variables in the environment record pointed to by $e$. As we have
already noted, the offset for each of these variables relative to the 
old environment record can be statically determined and so the
necessary initialization process can itself be compiled.

A second aspect to which special attention must be paid in the
presence of implication goals is the context switching necessitated by
backtracking. The broad requirement upon backtracking is to reinstate
a program that was in existence at some earlier point in the
computation. To understand clearly the changes that must be effected,
and consequently the bookkeeping that must be done, let us consider a
program containing the clauses

\begin{figure}
\begin{center}
\setlength{\unitlength}{0.0125in}
\begingroup\makeatletter\ifx\SetFigFont\undefined
\def\x#1#2#3#4#5#6#7\relax{\def\x{#1#2#3#4#5#6}}%
\expandafter\x\fmtname xxxxxx\relax \def\y{splain}%
\ifx\x\y   
\gdef\SetFigFont#1#2#3{%
  \ifnum #1<17\tiny\else \ifnum #1<20\small\else
  \ifnum #1<24\normalsize\else \ifnum #1<29\large\else
  \ifnum #1<34\Large\else \ifnum #1<41\LARGE\else
     \huge\fi\fi\fi\fi\fi\fi
  \csname #3\endcsname}%
\else
\gdef\SetFigFont#1#2#3{\begingroup
  \count@#1\relax \ifnum 25<\count@\count@25\fi
  \def\x{\endgroup\@setsize\SetFigFont{#2pt}}%
  \expandafter\x
    \csname \romannumeral\the\count@ pt\expandafter\endcsname
    \csname @\romannumeral\the\count@ pt\endcsname
  \csname #3\endcsname}%
\fi
\fi\endgroup
\begin{picture}(338,235)(0,-10)
\put(70,158){\ellipse{14}{14}}
\put(70,158){\blacken\ellipse{8}{8}}
\put(70,158){\ellipse{8}{8}}
\drawline(70,158)(70,158)
\put(110,98){\ellipse{14}{14}}
\put(110,98){\blacken\ellipse{8}{8}}
\put(110,98){\ellipse{8}{8}}
\drawline(110,98)(110,98)
\put(230,158){\ellipse{14}{14}}
\put(230,158){\blacken\ellipse{8}{8}}
\put(230,158){\ellipse{8}{8}}
\put(190,98){\ellipse{14}{14}}
\put(190,98){\blacken\ellipse{8}{8}}
\put(190,98){\ellipse{8}{8}}
\drawline(190,98)(190,98)
\put(150,198){\blacken\ellipse{8}{8}}
\put(150,198){\ellipse{8}{8}}
\put(30,98){\blacken\ellipse{8}{8}}
\put(30,98){\ellipse{8}{8}}
\put(270,98){\blacken\ellipse{8}{8}}
\put(270,98){\ellipse{8}{8}}
\drawline(250,153)(250,153)
\drawline(250,153)(250,153)
\path(150,198)(110,178)
\path(116.261,183.367)(110.000,178.000)(118.050,179.789)
\path(150,198)(190,178)
\path(181.950,179.789)(190.000,178.000)(183.739,183.367)
\path(190,178)(230,158)
\path(110,178)(70,158)
\path(70,158)(50,128)
\path(52.774,135.766)(50.000,128.000)(56.102,133.547)
\path(50,128)(30,98)
\path(70,158)(90,128)
\path(83.898,133.547)(90.000,128.000)(87.226,135.766)
\drawline(90,128)(90,128)
\path(90,128)(110,98)
\drawline(230,158)(230,158)
\path(230,158)(210,128)
\path(212.774,135.766)(210.000,128.000)(216.102,133.547)
\path(210,128)(190,98)
\path(230,158)(250,128)
\path(243.898,133.547)(250.000,128.000)(247.226,135.766)
\drawline(250,128)(250,128)
\path(250,128)(270,98)
\path(190,28)(190,58)
\path(192.000,50.000)(190.000,58.000)(188.000,50.000)
\path(270,28)(270,58)
\path(272.000,50.000)(270.000,58.000)(268.000,50.000)
\put(80,93){\makebox(0,0)[lb]{\smash{{{\SetFigFont{11}{13.2}{rm}$D_4$}}}}}
\put(125,93){\makebox(0,0)[lb]{\smash{{{\SetFigFont{11}{13.2}{rm}$p_4$}}}}}
\put(160,93){\makebox(0,0)[lb]{\smash{{{\SetFigFont{11}{13.2}{rm}$D_5$}}}}}
\put(205,93){\makebox(0,0)[lb]{\smash{{{\SetFigFont{11}{13.2}{rm}$p_5$}}}}}
\put(240,93){\makebox(0,0)[lb]{\smash{{{\SetFigFont{11}{13.2}{rm}$D_6$}}}}}
\put(285,93){\makebox(0,0)[lb]{\smash{{{\SetFigFont{11}{13.2}{rm}$p_6$}}}}}
\put(0,93){\makebox(0,0)[lb]{\smash{{{\SetFigFont{11}{13.2}{rm}$D_3$}}}}}
\put(45,93){\makebox(0,0)[lb]{\smash{{{\SetFigFont{11}{13.2}{rm}$p_3$}}}}}
\put(40,153){\makebox(0,0)[lb]{\smash{{{\SetFigFont{11}{13.2}{rm}$D_1$}}}}}
\put(85,153){\makebox(0,0)[lb]{\smash{{{\SetFigFont{11}{13.2}{rm}$p_1$}}}}}
\put(200,153){\makebox(0,0)[lb]{\smash{{{\SetFigFont{11}{13.2}{rm}$D_2$}}}}}
\put(245,153){\makebox(0,0)[lb]{\smash{{{\SetFigFont{11}{13.2}{rm}$p_2$}}}}}
\put(150,208){\makebox(0,0)[lb]{\smash{{{\SetFigFont{11}{13.2}{rm}$p$}}}}}
\put(225,133){\makebox(0,0)[lb]{\smash{{{\SetFigFont{11}{13.2}{rm}(1)}}}}}
\put(185,73){\makebox(0,0)[lb]{\smash{{{\SetFigFont{11}{13.2}{rm}(2)}}}}}
\put(90,13){\makebox(0,0)[lb]{\smash{{{\SetFigFont{11}{13.2}{rm}Implication goal prior to}}}}}
\put(90,0){\makebox(0,0)[lb]{\smash{{{\SetFigFont{11}{13.2}{rm}most recent choice point}}}}}
\put(250,13){\makebox(0,0)[lb]{\smash{{{\SetFigFont{11}{13.2}{rm}Most recent}}}}}
\put(250,0){\makebox(0,0)[lb]{\smash{{{\SetFigFont{11}{13.2}{rm}implication goal}}}}}
\end{picture}

\end{center}
\caption{Example of an Implication Tree}
\label{fig:imptree}
\end{figure}
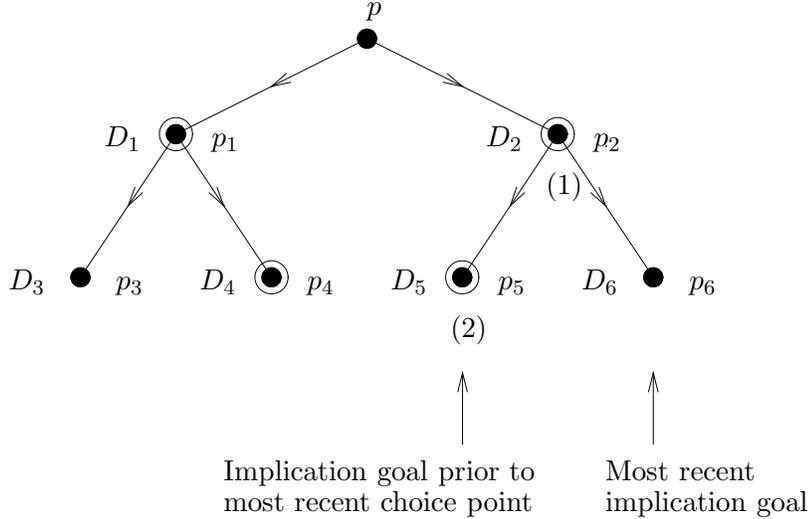

\begin{exmple}
\> $((D_1 \supset p_1) \land (D_2 \supset p_2)) \supset p$, \\
\> $((D_3 \supset p_3) \land (D_4 \supset p_4)) \supset p_1$,\\
\> $((D_5 \supset p_5) \land (D_6 \supset p_6)) \supset p_2$
\end{exmple}

\noindent and possibly others defining the predicates $p_3$,
$p_4$, $p_5$ and $p_6$; we assume that $D_1,\ldots, D_6$ represent
program clauses and that $p,p_1,\ldots, p_6$ are predicate names.
Suppose now that an attempt is made to solve the goal $p$. This
attempt engenders the invocation of implication goals whose dynamic
nature can be represented by a tree-like structure that we call an
{\it implication tree}. For instance, let us assume that the first
clause above is being used in the attempt to solve $p$, that, in this
context, a solution to $D_1\supset p_1$ has been found by using the
second clause to solve $p_1$ and that an attempt is now being made to
solve $p_2$ after having augmented the program with $D_2$. Let us
further assume that the third clause is being used in the attempt to
solve $p_2$, that $D_5 \supset p_5$ has already been solved in this
attempt and that a solution for $p_6$ is being sought from a program
that additionally contains the clause $D_6$. The state of the
computation as it relates to the invocation of implication goals can
then be depicted by the tree shown in Figure~\ref{fig:imptree}. The
nodes of this tree, with the exception of the root, correspond to the
invocations of implication goals, the clause that is added as a result
of this invocation being shown  to the left of the node and the goal
that is subsequently invoked being shown to the right. The root of
the implication tree represents the original goal. The left-to-right
ordering of the nodes reflects the time order of subcomputations and
the circles that are drawn around some nodes indicate the existence of
choice points subsequent to the goal invocation that they represent.

Suppose now that, in the situation being considered, the attempt to
solve the goal $p_6$ fails. The next step in the computation must then
be an attempt to find an alternative solution for the most recent
prior goal for which such a possibility exists. Referring to the
implication tree shown in Figure~\ref{fig:imptree}, this means that a
goal that lies below the node labelled (2) must be returned to. The
attempt to find another solution to this goal must, of course, be
preceded by a reinstatement of the program that was in existence when
the first (successful) attempt to solve it was made. This earlier
program state can be easily recreated if the implication tree is
available: We find, first of all, the closest common ancestor in the
implication tree of the node representing the most recent implication
goal, \ie, the latest implication goal below which the failure has
occurred, and the node denoting the last implication goal below which
the most recent choice point exists. Referring again to
Figure~\ref{fig:imptree}, this closest common ancestor is the node
labelled (1). The desired program is then obtained by discarding all
the program clauses that were added along the path from this node up
to and including the node representing the most recent implication
goal, and adding back all the program clauses that had been added (and
subsequently discarded) along the path from this node up to and
including the node denoting the last implication goal prior to the
most recent choice point.  In the example being considered, this
translates into discarding the program clause $D_6$ from the program
and adding back $D_5$.

Implementing the context switching process described above, requires a
record of the (annotated) implication tree and the nodes in this tree
representing the most recent implication goal in a global sense and
relative to each choice point to be maintained at each stage of
computation. We propose maintaining the implication tree by creating
an {\em implication point} record on the local stack at the start of
the computation and each time an implication goal is invoked. In order
to describe the structure of this record, we need to be more concrete
about the representation of the programs that are in existence at
various points in computation.  Recall that we intend to maintain
these as lists of closures. We shall assume for the moment that new
closures are added at the end of this list. The starting point of this
list is, therefore, fixed and the program available at any stage is
specified completely by the end point. We add a register named {\tt
LC} to the usual WAM registers for recording this end point relative
to any point in computation.  Taking this representation of programs
into account, an implication point record will have the following
information fields:

\begin{romanlist}

\item a reference, {\tt IC}, to the statically constructed table of
pointers to code for the clauses that constitute the antecedent of the
implication goal, 

\item a pointer, {\tt E'}, to the environment record that is current 
at the invocation of the implication goal that the implication point
record corresponds to,

\item a pointer, {\tt IP}, to the implication point record
for the most recent implication goal within which the
implication goal corresponding to the implication point record in
question appears embedded, and 

\item a pointer, {\tt LCP}, to the end of the list of closures
constituting the program at the time the implication goal is invoked.

\end{romanlist}

\noindent
Note that the start of the computation can itself be viewed as the
invocation of an implication goal whose antecedent is the original
program and whose consequent is the corresponding goal and the first
implication point record will be set up consistent with this viewpoint
and with its {\tt IP} field indicating that there is no enclosing
implication goal invocation.  Now, the nodes in the implication tree
are obviously represented by implication point records. Thus, the
remaining information that must be maintained for context switching
purposes consists of the most recent implication point record relative
to the current point in computation and to each choice point. The
former is maintained in another new register that is named {\tt I}.
For the latter, we add a field called {\tt IP} to the choice point
record of the WAM; this field will be set to the value of the {\tt I}
register at the time the choice point record is created.

We can now describe a scheme that accounts completely for implication
goals. Under this scheme, the invocation of an implication goal causes
an implication point record to be created and additions to be made to
the existing program. The augmentation to the program is carried out
in an obvious way using the {\tt E} register of the WAM and the table
generated at compile-time for the implication goal. The {\tt IC} field
of the implication point record is set simply to point to this table
and the contents of the {\tt E}, {\tt I} and {\tt LC} registers prior
to the invocation of the implication goal determine the {\tt E'}, {\tt
IP} and {\tt LCP} fields. At the end of this process, the {\tt I}
register is updated to point to the newly created implication point
record. Note also that the {\tt LC} register will be affected by the
augmentation to the program. When an implication goal is successfully
completed, the {\tt I} and {\tt LC} registers must be reset to their
values prior to the invocation of the goal. This is done by using the
relevant fields from the implication point record corresponding to the
goal that will be given by the contents of the {\tt I} register. The
implication point record can also be discarded if the implication goal
is more recent than the most recent choice point; this can be
determined simply by comparing the {\tt I} and the {\tt B} register
(that indicates the location in the stack of the most recent choice
point record as in the WAM) and, thus, the top of the local stack will
be given by the largest of the addresses in the {\tt I}, {\tt E} and
{\tt B} registers. Finally, suppose there is a failure at some point
in the computation. Assuming that there is an alternative solution
path to be explored, the appropriate program context is recreated as
follows: We 
first check if the {\tt I} register points to a location earlier on
the stack than the one pointed to by the {\tt B} register. If so, the
program context is already the appropriate one. Otherwise, we chain
back through implication point records starting from the record
pointed to by the {\tt I} register till we reach one that appears
lower in the stack than the location pointed to by the {\tt B}
register. Let us refer to this implication point record as {\it CCA}
(for closest common ancestor). We then discard the necessary closures
by setting the {\tt LC} register to the value of the {\tt LCP} field
stored in the implication point record just prior to {\it CCA} on the
path to it from the record pointed to by the {\tt I} register. The
{\tt I} register is then set from the {\tt IP} field of the most
recent choice point record. Finally, the addition of the relevant
closures is affected by using the {\tt IC} and {\tt E'} fields of the
implication point records between {\it CCA} and that given by the {\tt
I} register. The {\tt LC} register must, of course, be updated at the
end of this process.

We have assumed above that the addition of program clauses that
results from invoking an implication goal takes place at the end of
the program. Where exactly the addition should occur is unspecified by
the semantics that we have presented for the language. It is
conceivable, perhaps even desirable, that this addition be recorded at
the beginning of the program, thereby making the new clauses
accessible before the ones already in the program. The model described
above is amenable to this interpretation; it is the beginning of the
list of closures that must be recorded in this case instead of the
end.

\section{AN EFFICIENT REALIZATION OF PROGRAM CONTEXTS}\label{sec:rep}

The list representation of programs adopted in the last section is
useful from the perspective of understanding the addition and deletion
of program clauses. However, it is not adequate from a practical
standpoint since it does not allow rapid access to program clauses:
the list must be searched sequentially to determine the applicable
clauses. We can minimize the cost of this search by maintaining a
hash-table that is accessed by the name of the predicate. Each entry
in the hash-table points to a list of lists, one list for each
distinct predicate name that hashes to the same table entry. The list
associated with each predicate contains a closure for each program
clause that may be used in solving an atomic goal whose name matches
the predicate. Additions to a list may occur either at the end of this
list or at the beginning, depending on the chosen semantics.  In lieu
of the pointer into the original list of program clauses kept in each
implication point record, we must now maintain pointers into the
closure lists for {\it each} predicate that is defined in the
antecedent of the implication goal. These pointers facilitate the
discarding and reintroduction of closure entries upon successful
completion of an implication goal and upon backtracking. If the
antecedent of an implication goal contains multiple clauses for a
predicate, this scheme can be modified to accommodate the compilation
of sequencing with respect to these clauses.

In the case when new clauses are added to the front of a program, this
general idea can be implemented in a fashion that enables the context
switching required on backtracking to be realized with a minimum of
effort. The starting point for this scheme is an organization of the
global program, \ie, the program in existence prior to the user's
query, in a form that supports rapid access to the code for any given
predicate. In particular, we assume that multiple clauses for a
predicate give rise to one procedure with several entry points as in
the WAM and that this code may be located, for instance, by hashing on
the name of the predicate.  Now each time an implication goal is
invoked, new clauses may be added to the program for any given
predicate.  In order to provide efficient support for the process of
chaining through the clauses for a given predicate that are introduced
by different implication goals, we construct an access vector of
pointers that effectively identify the next most recent set of clauses
for the predicates defined by the clauses in the antecedent of the
implication goal.  This access vector is computed at the time the
implication goal is invoked and is saved in the implication point
record that is created for the invocation.  This implication point
record will be retained even after a solution to the corresponding
goal has been found so long as backtracking may cause the goal to be
retried.  Thus the access vector will be available too, and the old
program context can be resurrected simply by reverting to its use in
finding clauses for solving goals.

In spelling out the details of the scheme outlined above, we make use
of the discussions of the previous section. The central point, as
before, is the treatment of implication goals that appear in the
program. Once again, let
\[(D_1(\bar x_1)  \land \ldots \land D_n(\bar x_n)) ~\supset~ G\]
be a schematic representation of such a goal. Now, each of the clauses
$D_i(\bar x_i)$ will be compiled in the manner described in the
previous section. In contrast to the earlier situation, however,
clauses that define the same predicate will be combined into {\em one}
procedure with the use of clause sequencing code.  In general, this
will give rise to $m$ segments of code, defining $m$ predicates.
In conjunction with the implication goal, a table will be
created at compile-time with the following entries:

\begin{romanlist}

\item The number of predicates defined in the antecedent of the
implication goal, \ie, $m$ in the case considered.  We call this the
{\em size} of the implication goal. 

\item 
A pointer to code that, given any predicate name, either determines
that it is not defined by any of the clauses in the antecedent of the
implication goal or returns the location of the relevant compiled
code. The structure of the code that carries out this task will depend
on the number of predicates that are defined in the antecedent: if
this is a small number, then sequential search will suffice;
otherwise, a hash-table may be used.

\item A one-to-one mapping from the names of the predicates defined in
the antecedent of the implication goal to $\{1,\ldots,m\}$.  We refer
to the number associated with a particular name as its {\it offset
number} relative to the implication goal. This mapping is needed in
setting up implication point records as we explain presently.

\end{romanlist}

As mentioned above, access to program clauses will be provided through
implication point records in the new scheme.  This access is realized,
at a conceptual level, as follows. The search for clauses defining a
particular predicate takes place relative to an implication point
record that is pointed to by a new register called the {\tt CI}
register. At the outset, \ie, when an atomic goal is encountered, this
register is set to point to the most recent implication point record
by copying into it the contents of the {\tt I} register. If {\tt CI}
points to an implication point record representing the invocation of
an implication goal whose antecedent does not contain a clause
defining the predicate in question, then {\tt CI} will be updated to
point to the implication point record for the closest dynamically
enclosing implication goal invocation and the search will continue
from there. If there is no such enclosing implication goal, a failure
will result. (As before, we view the start of computation as the first
invocation of an implication goal, one for which an implication point
record is created at the bottom of the stack.)  On the other hand, if
there are clauses defining the predicate in the antecedent of the
relevant implication goal, then these will be used in an attempt to
solve the atomic goal. The use of these clauses will, in general,
require bindings for certain variables to be initialized from an
appropriate environment record. The location of this environment
record is available from the implication point record pointed to by
{\tt CI} and will be copied into another new register called {\tt CE}
before the clauses in question are used.

Now suppose that all the clauses available for a predicate through a
particular implication point record have been tried and have resulted
in failure. There may still be clauses available that can be tried in
an attempt to solve the given atomic goal. Conceptually these clauses
can be located by chaining back through implication point records for
the enclosing implication goals. However, this work can be reduced by
doing it once and for all at the time the implication point record is
created. In particular, for each predicate defined in the relevant
implication goal, we can compute and store a pointer to the closest
enclosing implication point record containing a clause for that
predicate and a pointer to the corresponding code.  This optimization,
while useful in general, turns out to be particularly helpful in
compiling clause sequencing as we shall see in the next section.

Taking the various discussions of this section into account, the
information that is now to be stored in an implication point record 
is the following: 

\begin{alphabetlist}

\item A pointer, {\tt IC}, to the code for determining whether a
predicate is defined by any clauses in the antecedent of the
corresponding implication goal and, if it is, for finding the address
of the code generated from these clauses.  The address to be stored in
{\tt IC} is available from the table compiled for the implication
goal; see item (ii) above.  (This field is conceptually similar to one
of the same name in the implication point record of
Section~\ref{sec:impl}.)

\item A pointer, {\tt E'}, to the environment record that is current
at the invocation of the implication goal that the implication point
record represents.

\item A pointer, {\tt IP}, to the implication point record
for the most recent implication goal within which the implication goal
corresponding to the implication point record in question appears
embedded.

\item An access vector, $nc$, whose size is that of the implication
goal. The $i$th entry of this vector contains a pointer to the code
for the next clause for the predicate with offset number $i$ and a
pointer to the implication point record in which this clause
``occurs''; if such a clause does not exist, the address of a failing
procedure is inserted.

\end{alphabetlist}

\noindent The last component is computed at the time the implication
point record is created.\footnote{An alternative approach is possible:
the access vector may be computed ``on demand'', \ie, the location of
the next clause may be computed when first needed and stored in the
vector to facilitate a direct lookup on subsequent occasions.}\ The
manner in which this computation is carried out should be obvious from 
the previous comments.

The availability and interpretation of the code for clauses within the
scheme outlined is dependent on the values of the {\tt CI} and {\tt
CE} register.  Consequently these registers must be saved in the
choice point record.  Accordingly, our choice point records contain
three new fields in comparison with the WAM, the {\tt IP}, the {\tt
CIP} and the {\tt CEP} fields. We also observe the ease with which the 
program context can be reset to the required value upon backtracking:
the current program and the clauses yet to be tried in solving a
particular atomic goal are determined by the value of the {\tt I} and
{\tt CI} and {\tt CE} registers respectively, and it is only necessary
to set these registers from the corresponding fields in the most
recent choice point record.

It is useful to understand qualitatively the cost of the proposed
scheme for supporting a scoping ability relative to program clauses.
At the very outset, we note that merely having the flexibility of
changing the program context dynamically incurs an overhead even if it
is not used, \ie, even if the program consists solely of program
clauses from the Horn clause setting.
There are two sources for this overhead. First, each choice point
record must store three extra fields --- the {\tt IP}, {\tt CIP} and
{\tt CEP} fields --- with associated space and time costs. Second, the
location for the code to be used in solving a given atomic goal can
only be determined dynamically, perhaps via a hash-table. Allowing for
universal goals adds one more field, the {\tt UIP} field, to choice
point records and incurs a cost for tagged unification whose precise
nature will become evident in the next section.

Certain costs are incurred in addition to those above if a genuine use
is made of the scoping ability relative to program clauses that is
afforded by our language. First of all, the search for the code for a
predicate becomes more complex.  A reasonable assumption for the time
required to locate code for a predicate in a program unit, \ie, the
block of code corresponding to the original program or the antecedent
of an implication goal, is that it is fixed.  Viewing the top-level
goal itself as an implication goal, the (time) degradation in locating
the code for a predicate then depends on the deviation from $1$ of the
number of nested implication goal invocations within which the attempt
to find such code takes place.  In assessing the overall degradation,
it is necessary to amortize the number of nested implication goal
invocations over all procedure calls and also to consider the
proportion of all operations that procedure calls constitute.  Taking
these aspects into account and noting that well-written programs
should result in only a small nesting of implication goals, we believe
that the overhead due to this factor will be small.  A second factor
affecting performance is the need to set up and maintain implication
point records.  Let $n$ be the number of predicates that are defined
in the antecedent of the implication goal corresponding to an
implication point record.  The space required for the record is then
$3+2*n$ pointers. The only time expenditure in creating the record
that is not fixed is that for setting up the access vector.  As
already noted, the time needed for locating the code for a predicate
in a dynamic context is proportional to the number of nested
implication goal invocations by which the context is defined. The time
required for computing the access vector would be $n$ times this cost.
The number of predicates defined in the antecedent of an implication
goal and the nesting level of implication goal invocations will, in
the typical situation, be bounded by a small number.  The overall
space and time costs due to this factor will thus be roughly
proportional to the number of implication point records that are set
up in the course of solving a query. This number can be assumed to be
small, especially in comparison with the number of procedure calls and
other operations that will have to be performed.

Before concluding this section, we note the similarity between the
scheme that we have outlined here and that used for contextual logic
programming in \cite{LMN89}. Indeed, the mechanisms presented in this
section are an amalgamation of the ideas discussed in Section
\ref{sec:impl} (and in \cite{JN91}) and those in \cite{LMN89}. Our
scheme differs in detail from that in \cite{LMN89} in that (a) we
eliminate the context stack by using implication point records that
are stored on the local stack, (b) we need to deal with closures
instead of just program code, and (c) implication goals involve only
one of the several semantics that are implemented in
\cite{LMN89}.\footnote{An implication goal $D \supset G$ can be
interpreted as the goal $U >> {\bar G}$ in contextual logic
programming where $U$ is a unit containing a translated version 
of the program clauses in $D$ 
and ${\bar G}$ is the translation of $G$ obtained by using this
transformation recursively. Under this interpretation, the 
operational semantics we have defined for implication goals 
corresponds in contextual logic programming to assuming that 
the clauses in a unit extend the definitions of predicates 
available in a dynamic context and that goals are solved 
by using a lazy binding in the sense of \cite{LMN89}.}

\section{COMPILATION}\label{sec:compile}

A scheme for compiling a logic programming language into WAM-like
instructions must address two main issues: the compilation of
unification and the compilation of control.  The same general approach
can be used in a treatment of these aspects in the context of our
language as in the case of Prolog.  However, there are differences in
detail, arising from the fact that some new problems have to be
handled in an implementation of our language.
We have presented schemes for dealing with these problems in earlier
sections and have also indicated the possibility of compilation within
these schemes. We provide concreteness to the latter discussion in
this section by describing modifications and additions to the
instructions of the WAM for accounting for (a)~the tagged form of
unification, (b)~the larger variety of non-atomic goals, and (c)~the
possibility that the clauses that appear in the antecedents of
implication goals actually extend previously existing definitions of
predicates. We also illustrate the use of the resulting instruction
set in compiling programs in our language.

\subsection{Compilation of Unification}\label{ssec:compunif}
We shall assume that the set of instructions that are included in the
WAM for the purpose of compiling unification is that described in
\cite{HAK91} as opposed to the one contained in \cite{War83}. The main
difference between these two sets is that the former includes a
collection of {\tt set} instructions that parallel the {\tt unify}
instructions. These {\tt set} instructions are used instead of the
{\tt unify} instructions in compiling the creation of terms in the
scope of the {\tt put\_structure} and {\tt put\_list}
instructions. While this ``enhancement'' to the instruction set is not 
essential, it is useful in reducing mode setting and testing 
in the context of the WAM and also provides the basis for avoiding
some occurs-checking and the checking of tag compatibility in our
context. Now, despite the changed nature of unification for our
language, no instructions are needed in addition to those already in
the WAM for implementing this operation. However, some of the WAM
instructions must be modified to ensure that tags are maintained and
respected during unification.

The tagging of variables is dependent on their classification as
either temporary or permanent.  This classification must be performed
relative to each program clause whose compilation is to be considered,
\ie, relative to each clause that is part of the original program or
that appears as one of the conjuncts in the antecedent of an
implication goal.  The variables of such a clause that need to be
classified as temporary or permanent are the following: (a)~those that
are free in the clause --- this is relevant only in the case that the
clause appears in the antecedent of an implication goal, (b)~those
that are (implicitly) universally quantified over the clause, and
(c)~those that are 
explicitly quantified in the body of the clause but where the
quantification is not embedded in the antecedent of an implication
goal.  Of these variables, those that are universally quantified in
the body of the clause or have an occurrence in the antecedent of an
implication goal appearing there are considered permanent.  An
existentially quantified variable or a variable of category (b) is
also considered permanent if it has an occurrence in a universal goal
that appears within the scope of the quantifier governing the
variable.  The categorization of the remaining variables is
determined after the given clause is reduced to one in the Horn clause
setting by dropping quantifiers in its body and replacing implication
goals by their consequents.  Free variables, \ie, variables of
category (a), are considered temporary if their occurrences are
limited to the head and first goal in the body under this reduction
and permanent otherwise.  Finally, for the rest of the variables we
use the classification employed with the WAM, with the proviso that a
goal that originally appeared embedded inside an implication or a
universal quantifier is not to be considered a last goal under the
reduction.

\ignore{This classification is somewhat different in our context
from that used in the context of the WAM. The only variables that need
to be classified in this fashion with respect to a program clause are
(a)~those that are free in the clause, (b)~those that are (implicitly)
universally quantified over the clause, and (c)~those that are
explicitly quantified in the body of the clause but where the
quantification is not embedded in the antecedent of an implication
goal. Those of these variables that have an occurrence in the
antecedent of an implication goal in the body of the clause are
considered permanent. A variable of category (a) or (b) is also
considered permanent if its first occurrence is within the scope of a
universal goal. An existentially quantified variable of kind (c) is
similarly considered permanent if its first occurrence is in the scope
of a universally quantified goal that is itself within the scope of
the quantifier for the variable. The categorization of the remaining
variables is determined by first reducing the clause to a Horn clause
by dropping quantifiers in its body and replacing implication goals by
their consequents and then using the scheme employed in the context of
the WAM.} 

No tags are associated with temporary 
variables initially; tags for these variables are determined by the
instructions that manipulate them as we see below.
The permanent variables of a clause are tagged with the value of the
universe index at the time the clause is invoked. 
The relevant tag value is obtained from the {\tt UI} register and the
tagging action is carried out by the {\tt allocate} instruction. This
instruction is, in our case, provided with an argument indicating a 
number of suitably tagged unbound references that are to be created on
the top of the stack.
This action makes unnecessary the initialization that is 
performed by {\tt put\_variable} and {\tt set\_variable} relative to
permanent variables. These instructions can therefore be eliminated
and the {\tt put\_value} and {\tt set\_value} instructions can be used
in their place. 

The unification related instructions are changed in the following
fashion. 
The instructions that write constants must now also associate the tag
$1$ with these constants.   
This requirement affects the instructions {\tt put\_constant}, {\tt
set\_constant} and, in the appropriate contexts, {\tt get\_constant}
and {\tt unify\_constant}.  
Instructions that bind or create variable cells must, similarly, be
sensitive to tag associations. 
Among these, it turns out that the instructions {\tt get\_variable}
and {\tt unify\_variable} (and {\tt unify\_void}) executed in {\it
read mode}, do not need to handle tags at all; the binding will always
be permitted and the incoming structure, variable or constant will
carry the necessary tags.  
The {\tt set\_variable} and {\tt set\_void} instructions must tag the
variable cells that they create on the heap with the value of the
{\tt UI} register. 
The {\tt put\_variable} instruction, used now only with respect to a
temporary variable, must perform a similar association. 
The instructions {\tt put\_unsafe\_value} and {\tt set\_local\_value}
create new variable cells on the heap in
certain situations and, in these cases, they must associate the tag
value of the stack variable that is being ``copied'' with the newly
created cell. 
Finally, when the {\tt unify\_variable} and {\tt unify\_void}
instructions are executed in {\it write mode}, they must associate a
tag value with the variables being written that is equal to the tag
value of the variable whose value is being set by the governing {\tt
get\_structure} instruction.\footnote{The comments of Pascal Brisset
made us aware of 
an error with regard to this point in an earlier version of our
implementation scheme. The same observation was also made by one of
the authors of this paper, Keehang Kwon.}\ To facilitate the
communication of this tag value between the {\tt get\_structure} and
the {\tt unify\_variable} instructions, use is made of a new register
called the {\tt UT} register. The {\tt get\_structure} instruction
copies the relevant tag value into this register when it encounters an
incoming argument that is an unbound variable. 

The instructions considered up to this point only require
modifications to ensure that the right tag values are written with
variables and constants. The only times at which the compatibility of
tags need to be checked are when two constants are being matched by
{\tt get\_constant} or {\tt unify\_constant} and within the
unification process that is carried out in interpretive mode in
conjunction with {\tt get\_value}, {\tt unify\_value} or {\tt
unify\_local\_value}. In the former case, the check that must be made 
amounts simply to considering tag values to be parts of the
names of constants. In the latter case, the necessary check causes
variable assignments to be constrained according to the following: A
variable cannot be bound to a constant with a higher tag or to a
structured term containing a constant with a higher tag. If a variable
is bound to another variable with a higher tag or to a structured term
containing a variable with a 
higher tag, then the tag value of the latter variable must be set to
that of the former. A {\tt unify\_value} or {\tt unify\_local\_value}
instruction executed in {\it write mode} is already a part of a
variable assignment. The tag value 
of the variable being assigned to is contained in the {\tt UT}
register and it is this value that must be used in the described check
of tag compatibility. Note that the interpretive unification
process that is being considered must
include an occurs-check in an implementation that is sound with
respect to the logic considered in this paper or, for that matter,
with respect to Horn clause logic. The additional checking of tag
compatibility is similarly needed for soundness in the case of our
language and can, in fact, be carried out in the same phase as the
occurs-check. There is a possibility of avoiding both the occurs-check 
and the checking of tag compatibility in certain situations and doing
so may well be important to the efficiency of an actual
implementation. However, a detailed examination of this issue is
beyond the scope of this paper.

\subsection{Compiling Complex Goals}

The issue of concern here is the compilation of the logical symbols
that may appear in goals. The symbols in question are $\lor$, $\land$,
$\all$, $\exists$ and $\supset$. Goals of the form $G_1 \lor G_2$ and
$G_1 \land G_2$ are also permitted in Prolog and the method of
treatment used there is adequate in our context as well. In
particular, $\lor$ gives rise to code for generating a choice point
record and $\land$ results in the sequential execution of the code for
the subgoals.

The treatment of the universal quantifier follows the lines indicated
in Section~\ref{sec:interp}. Thus, consider the goal $\allx{x}G$.
Bearing in mind the classification of variables described in
Subsection~\ref{ssec:compunif}, the variable $x$ would be deemed a
permanent variable in the context in which this goal is encountered
and so a cell will be allocated for it in the current environment
record. Now, the code that is generated for the given goal must
increment the {\tt UI} register, place a new constant whose tag value
is that contained in the {\tt UI} register in the cell allocated for
$x$ and then execute the code for $G$. Further, if the code for $G$
completes successfully, the {\tt UI} register must be decremented.
Three new instructions are introduced for supporting these
requirements:
\begin{exmple}
\> {\tt incr\_universe}\\
\> {\tt decr\_universe}\\
\> {\tt set\_univ\_tag Yi}
\end{exmple}
\noindent The first two instructions respectively increment and
decrement the {\tt UI} register, and the last
instruction binds the (permanent) variable {\tt Yi} to a
new constant that is tagged with the value of the {\tt UI} register. 

A final comment concerning the treatment of universal goals is that,
as noted in Section~\ref{sec:interp}, the value of the {\tt UI}
register must be stored in the {\tt UIP} field of a choice point
record at the time that this record is created.

The action to be performed in conjunction with an existentially
quantified goal depends on whether the quantified variable is
classified as permanent or temporary. Suppose that the goal
that is encountered is $\somex{x}G$. If $x$ is considered to be a
permanent variable, then the tag value of the cell allocated for $x$
must be set using the {\tt UI} register and the compiled code for $G$
must be executed. On the other hand, no tags need to be set if 
$x$ is considered to be a temporary variable and execution can
proceed directly to the code for $G$. In realizing these actions,
there is need for only one new instruction:
\begin{exmple}
\>{\tt set\_exist\_tag Yi}.
\end{exmple}
\noindent  This instruction tags the permanent variable {\tt Yi} with 
the value of the {\tt UI} register. 

\ignore{The action to be performed for existential quantifiers is
relatively straightforward. When a goal of the form $\somex{x}G$ is
encountered, the tag value of the permanent or temporary variable
associated with the bound variable must be made the value of the {\tt
UI} register and the compiled code for $G$ must be executed. One new
instruction is required for this purpose:  
\begin{exmple}
\>{\tt set\_exist\_tag Vi}
\end{exmple}
\noindent  This instruction tags the variable {\tt Vi} with the value
of the {\tt UI} register. 
}

The treatment of implication goals was discussed in detail in the
previous section. Recalling this, when a goal of the form $D\supset G$
is encountered, an implication point record representing the addition
of $D$ to the program must be pushed onto the local stack and access
to the resulting program must be relativized to this record. In the
case that the implication goal completes successfully, access to the
program must be restored to being through the implication point record
pointed to by the {\tt I} register prior to the invocation of the
goal. Compilation of these actions is supported by the following new
instructions:
\begin{exmple}
\>{\tt push\_impl\_point t,n} \\
\>{\tt pop\_impl\_point}
\end{exmple}
\noindent In the first instruction, {\tt t} represents a pointer to
the statically created table for an implication goal that was
described in Section~\ref{sec:rep} and {\tt n} represents the number
of variables in the current environment record. This instruction
results in an implication point record being pushed onto the top of
the local stack, this being located by examining the {\tt I} and {\tt
B} registers and the {\tt E} register plus the size of the current
environment record. The manner in which the instruction fills in the
fields of the implication point record should be obvious from the
discussions in the last section. After creating the implication point
record, the instruction updates the {\tt I} register to point to it.

The instruction {\tt pop\_impl\_point} simply restores the previous
value of the {\tt I} register by using the {\tt IP} field of the
implication point record that the register currently points to.

\subsection{Compiling Atomic Goals and Clause Sequencing}

The compilation of clause sequencing for the global program, \ie, the
program in existence prior to the user's query, remains unaltered from
that used for Prolog relative to the WAM. However, there is a slightly
different interpretation to the instructions that are used. Those
instructions that create a choice point record --- specifically, {\tt
try\_me\_else} and {\tt try} --- must now also store the contents of
the {\tt UI}, {\tt I}, {\tt CI} and {\tt CE} registers in the record.
Correspondingly, the backtracking action performed by the instructions
{\tt retry\_me\_else}, {\tt retry}, {\tt trust\_me} and {\tt trust}
must include a restoration of the values of the {\tt UI}, {\tt I} and
{\tt CE} registers from the relevant choice point record.

The clauses in the antecedent of an implication goal are compiled
assuming that they constitute a unit, distinct from the rest of the
program. The code that is generated for predicates defined in this
unit differs from that that would be generated in the case of the
global program in only two respects. The first difference is that the
code produced for those clauses that have free variables in them will
have a part that relativizes the bindings for these variables to the
current environment record. The new instruction
\begin{exmple}
\>{\tt initialize Vn,m}
\end{exmple}
\noindent in which {\tt Vn} is a temporary or a permanent variable
and {\tt m} is a number is used to achieve this effect.  This
instruction is like the {\tt get\_variable} instruction of the WAM
except that the second argument is obtained by using the
{\tt m}th variable from the environment record pointed to by the {\tt
CE} register. The second difference is that the code that is generated
will always contain the creation of a choice point record and its last
instruction will be one that has the effect of attempting other
clauses that may be available in the dynamic context for the relevant
predicate. The instruction
\begin{exmple}
\>{\tt trust\_ext Pi}
\end{exmple}
\noindent is added for this purpose.\footnote{We borrow this
instruction from \cite{LMN89}.}\ In this instruction, {\tt Pi} is an
offset number relative to an implication goal.  When the clauses for a
particular predicate that appear in the antecedent of an implication
goal are compiled, the code for the last clause is preceded by a {\tt
try\_me\_else Li} or a {\tt retry\_me\_else Li} instruction and is
followed by
\begin{exmple}
\> {\tt Li : trust\_ext Pi}
\end{exmple}
\noindent where {\tt Pi} is the offset number for the predicate.
Executing this instruction has the following effect: The current
choice point record is used to reset all the registers except {\tt P},
which is the program pointer as in the WAM.  The entry at location
{\tt Pi} in the $nc$ field of the implication point record pointed to
by {\tt CI} is then used to set the {\tt CI} and {\tt P} registers.
Finally, the {\tt CE} register is set to the {\tt E'} field of the
record pointed to by {\tt CI}.

With regard to the compilation of an atomic goal, the code for
preparing the argument registers follows the pattern used relative to
the WAM. The actual invocation of the code for the corresponding
predicate name is also achieved through the {\tt call} or {\tt
execute} instruction. However, these instructions have a different
interpretation in our case from that in the WAM.  For example,
consider {\tt call q,n}.  The search that is made for code for {\tt q}
in executing this instruction must depend on the dynamic context. This
search starts by setting {\tt CI} to the value in {\tt I} and proceeds
in the fashion outlined in the previous section. If code is found, it
is executed as described. Otherwise backtracking occurs.

\subsection{Examples of Compiled Code}

We adopt below the Prolog conventions of writing implications in
program clauses backwards and depicting it by the symbol {\tt :-}, of
representing conjunctions in clause bodies by commas and of leaving
the top-level universal quantifiers implicit. We present two examples,
one illustrating the compilation of multiple clauses in the antecedent
of an implication goal that define the same predicate, and the other
illustrating the processing of a mixture of quantifiers in goals.

For the first example, we use one of the definitions of {\tt rev} from
Section \ref{sec:examples}:
\begin{tabbing}
\hspace{2em}\=\hspace{2em}\={\tt ((}\=\hspace{5em}\=\kill
\> {\tt rev(L1,L2) :-}\\
\>\> {\tt ((rev\_aux([],L2) $\land$}\\
\>\>\> {\tt ($\all$X$\,\all$L1$\,\all$L3$\,$(rev\_aux([X|L1],L3) :-
rev\_aux(L1,[X|L3]))))} \\
\>\>\>\> $\supset$ {\tt rev\_aux(L1,[])).}
\end{tabbing}
\noindent This clause has one permanent variable, namely {\tt L2}.
This variable occurs in the first clause for {\tt rev\_aux} and the
code for that clause must include an instruction for initializing it.
We assume that the statically determined table for the implication
goal that appears in the definition of {\tt rev} is pointed to by {\tt
t1}. The compiled code corresponding to {\tt rev} is then the
following:

\smallskip

\begin{tabbing}
\quad\={\tt rev\_aux: }\={\tt push\_impl\_point t1,4}\quad\=\kill
\>{\tt rev:}\>{\tt allocate 1}\\
\>\>{\tt get\_variable Y1,A2 }\\
\>\>{\tt push\_impl\_point t1,1}\> {\tt \% add rev\_aux code}\\
\>\>{\tt put\_constant [],A2 }\\
\>\>{\tt call rev\_aux,1}\\
\>\>{\tt pop\_impl\_point}\>{\tt \% restore earlier program}\\
\>\>{\tt deallocate}\\
\>\>{\tt proceed}
\end{tabbing}

\smallskip

\noindent Note that the {\tt call} instruction is used here instead of
the {\tt execute} instruction for invoking the {\tt rev\_aux}
procedure. This invocation appears to be the last call in the body of
the clause and it may therefore seem that the code that is shown does
not include the last call optimization that is common to Prolog
implementations. However, a little thought reveals this not to be the
case. The last action that must be performed actually relates to the
implication goal that forms the body of the clause: the clauses that
are added in the course of solving it must be removed {\it after}
solving {\tt rev\_aux}. It might be possible to include this action
within the code produced for {\tt rev\_aux}, thereby permitting the
environment record for {\tt rev} to be discarded before this code is
invoked.  However, it seems unlikely that doing this will improve
space usage significantly. The environment record that is retained at
present only contains bindings for tied variables and continuation
information and any modified scheme will also have to maintain such
information. Furthermore, the main utility of last call optimization
is in the context of recursive calls and it is reasonable to assume
that such calls will not appear repeatedly in situations where the
program is being extended, \ie, embedded within implication goals. We
note in this connection that our scheme {\em does not} affect the
usual applicability of last call optimization.

The following code would be generated for the two clauses defining
{\tt rev\_aux} in the body of the implication goal:

\smallskip

\begin{tabbing}
\quad\={\tt rev\_aux: }\={\tt push\_impl\_point c1,4}\quad\=\kill
\>{\tt rev\_aux:}\>{\tt try\_me\_else C1}\\
\>\>{\tt initialize X3,1}\>{\tt \% X3 = L2}\\ %
\>\>{\tt get\_constant [],A1}\>\\
\>\>{\tt     get\_value X3,A2 }\>{\tt \% unify L2 and second
argument}\\ 
\>\>{\tt proceed}\\
\\
\>{\tt C1:}\>{\tt retry\_me\_else C2}\\
\>\>{\tt get\_list A1}\\
\>\>{\tt unify\_variable X3}\\
\>\>{\tt unify\_variable A1}\\
\>\>{\tt get\_variable X4,A2}\\
\>\>{\tt put\_list A2}\\
\>\>{\tt set\_value X3}\\
\>\>{\tt set\_local\_value X4}\\
\>\>{\tt execute rev\_aux}\\
\\
\>{\tt C2:}\>{\tt trust\_ext 1}
\end{tabbing}

\smallskip

\noindent A point to note with respect to this code is that the 
choice point record is not discarded before the second clause for {\tt 
rev\_aux} is used. The reason for this is that this clause may not be
the last one for the predicate in the relevant dynamic context. As
observed in Section~\ref{sec:examples}, a universal quantification
over {\tt rev\_aux} will ensure that this is the case and will, in
fact, provide this information to a compiler as well. Such a
quantification is not permitted in the language currently being
considered, but is included in the extension that we examine in the
next section.

The second example that we consider is that of compiling the clause 
\smallskip

\begin{tabbing}
\hspace{2em}\={\tt p(Y) :- }\={\tt
($\all$U$\,\some$Z$\,$}\=\hspace{2em}\=\kill 
\>{\tt p(Y) :- ($\all$U$\,\some$Z} \\
\>\>\>{\tt (($\all$W$\,$(d1(Y,W,Z) :- r(Y,W)))} $\land$\\
\>\>\>{\tt ~($\all$W$\,$(d2(Z,W) :- d1(Z,W,W)))}\\
\>\>\>\>{\tt $\supset$ $\some$V$\,$g(Z,U,Y,V)),}\\
\>\>{\tt  ~~h(Y).}
\end{tabbing}

\smallskip

\noindent
In generating code for this clause, it is necessary to determine the
free variables of the clauses that form the antecedent of the
implication goal that appears in its body. These variables are those
that appear in the relevant clauses and whose (implicit or explicit)
quantification governs the implication goal. Thus, the free variables
of the clause defining the predicate {\tt d1} are {\tt Z} (explicitly
quantified) and {\tt Y} (implicitly quantified) and the only free
variable of the clause defining {\tt d2} is {\tt Z}. Bindings for
these variables must be contained in the environment record
corresponding to {\tt p} at a point when the respective clauses are
invoked and it is for this reason that they are deemed permanent
variables of the clause defining {\tt p}. The variables {\tt U} and
{\tt V} are also permanent variables of this clause and, assuming that
{\tt t2} points to the table constructed for the implication goal, the 
code that would be generated for the clause is the following:

\smallskip

\begin{tabbing}
\quad\={\tt d2: }\={\tt push\_impl\_point c1,4}\quad\=\kill
\>{\tt p:}\>{\tt allocate 4}\\ %
\>\>{\tt get\_variable Y1,A1}\\ %
\>\>{\tt incr\_universe}\> {\tt \% ($\all$} \\
\>\>{\tt set\_univ\_tag Y2}\>{\tt\% \ \ U}\\
\>\>{\tt set\_exist\_tag Y3}\>{\tt \% \ \ \ $\some$Z}\\
\>\>{\tt push\_impl\_point t2,4}\>{\tt \% \ \ \ \ \ add clauses for d1 
and d2}\\ 
\>\>{\tt set\_exist\_tag Y4}\>{\tt \% \ \ \ \ \ \ $\some$V}\\
\>\>{\tt put\_value Y3,A1}\> {\tt \% \ \ \ \ \ \ \ \ g(Z,} \\
\>\>{\tt put\_value Y2,A2}\> {\tt \% \ \ \ \ \ \ \ \ \ \ U,} \\
\>\>{\tt put\_value Y1,A3}\> {\tt \% \ \ \ \ \ \ \ \ \ \ Y,} \\
\>\>{\tt put\_value Y4,A4}\> {\tt \% \ \ \ \ \ \ \ \ \ \ V} \\
\>\>{\tt call g,4}\> {\tt \% \ \ \ \ \ \ \ \ \ ),}\\
\>\>{\tt pop\_impl\_point}\>{\tt \% \ \ \ \ \ discard clauses for d1
and d2}\\ 
\>\>{\tt decr\_universe}\> {\tt \% ),} \\
\>\>{\tt put\_value Y1,A1}\> {\tt \% h(Y }\\
\>\>{\tt deallocate}\> {\tt \%}\\
\>\>{\tt execute h}\> {\tt \% \ )}
\end{tabbing}

\smallskip

We do not present the code for the clauses {\tt d1} and {\tt d2}. The
structure of this code should be clear from the previous example.

\section{DEALING WITH HIGHER-ORDER ASPECTS}\label{sec:higherorder} 

The propositional and quantifier structure of goals and program
clauses in the theory of higher-order hereditary Harrop formulas bears
a close similarity to that for these formulas in the first-order
language considered so far. One distinction is that, for reasons of
logical consistency, the higher-order formulas must be typed.  No new
implementation issues arise when a simple, non-polymorphic, form of
typing is used and we implicitly assume such a typing regimen below;
the treatment of polymorphic typing is considered in detail in
\cite{KNW92}. Another difference is that, for technical reasons, the
vocabulary of the higher-order logic includes the symbol $\top$ to
denote the tautologous proposition and this symbol is considered to 
be an acceptable goal. The final and most significant
difference is that first-order terms are replaced
by the terms of a (simply typed) lambda calculus. 

The lambda terms used in a higher-order logic can generally contain
arbitrary quantifiers and connectives in them. However, for reasons
explained in \cite{MNPS91}, our higher-order logic does not permit the
terms that it uses to contain the symbols $\supset$ and $\neg$. The
terms that result from omitting these symbols are referred to as {\it
positive} terms. A (positive) atomic formula is then a formula of the
form $P( t_1,\ldots,t_n)$ where $P$ is a predicate constant or
variable and, for $1 \leq i \leq n$, $t_i$ is a positive term.  Such a
formula is said to be {\it rigid} in the case that $P$ is a constant
and {\it flexible} otherwise. Using the symbol $A_r$ to represent
rigid atomic formulas and $A$ to denote arbitrary atomic formulas, the
higher-order versions of goals and program clauses are given by
the following syntax rules:
\[
\begin{tabular}{l}
$G ::= \top\sep A \sep (G\land G) \sep (G\lor G) \sep (\somex{x} G)
\sep (\dset \supset G) \sep (\allx{x} G)$,\\
$\dset ::= D \sep (D \land \dset)$,\ {\rm and}\\
$D ::= A_r \sep (G \supset A_r) \sep (\allx{x} D)$.\\
\end{tabular} \]

From an implementation perspective, the main new concern in
conjunction with our higher-order language is that first-order
unification must be replaced by a notion of unification that
incorporates equality based on $\lambda$-conversion.  The resulting
unification problem is different in several respects from first-order
unification. In particular, the problem is undecidable in general and
most general unifiers might not exist even when there are unifiers for
given terms. There is, nevertheless, a procedure that can be used to
find unifiers for these terms whenever they exist \cite{Huet75}. This
procedure can be factored into the repeated application of certain
simple steps and can be amalgamated as such into the abstract
interpreter described in Section \ref{sec:interp}. A similar
amalgamation has been carried out in \cite{NM90} relative to a
higher-order version of the Horn clause language and has been used in
\cite{NJW92} to describe a WAM-based implementation scheme for this
language. At a level of detail, the main new
implementation concerns in the context of this language are
(a)~devising a good 
representation for lambda terms, (b)~including machinery for
performing $\lambda$-conversion, (c)~incorporating a mechanism that
supports the explicit representation of sets of terms that have to be
unified, and (d)~handling the possibility of branching within
unification. The implementation scheme described in \cite{NJW92}
contains a treatment of all these aspects.  The language of interest
here, the one described by the $D$ and $G$ formulas above, results
essentially from adding universal quantifiers and implications as
scoping devices to the higher-order Horn clause language.  The
approach to implementing these scoping mechanisms that we have
presented in this paper carries over readily to the higher-order
context. No significant changes are necessary with regard to the
treatment of implication goals. The treatment of universal quantifiers
must take into account the fact that predicate and function symbols
can also be quantified over in the higher-order language.  Tags must
therefore be associated with these symbols as well and these must be
used in the course of unification.  An implementation of a
higher-order language must already countenance the fact that variables
and constants can be of function and predicate type and so the
association of tags can be carried out in a manner entirely consistent
with that described in this paper.  The use of tags can be described
as a simple check for tag compatibility at the time of binding a
variable even in the context of the higher-order language
\cite{Nad92int}, and this check can be implemented in a fairly
transparent fashion. 

A problem that is not directly addressed by the considerations above
is that of complex goals that are generated dynamically. In the
higher-order context, it is possible for a program to contain a goal
of the form $P(a)$ where $P$ is a variable. Now, $P$ might be
instantiated in the course of computation so that this goal becomes
one that, for instance, has a universal quantifier as its top-level
logical symbol. This raises the question of what code should be
produced for the goal $P(a)$ by the compilation process. Clearly, it
is not possible to anticipate the run-time form of this goal and so a
compiler cannot produce code that accords a direct treatment to this
form.  However, an indirect treatment that fits in well with our
current implementation scheme can be provided.  The essential idea is
to replace the goal $P(a)$ by the goal $solve(P(a))$, where $solve$ is
a predicate that is defined by the clauses (written in pseudo-Prolog
syntax)
\begin{exmple}
\>{\tt $solve(G_1 \land G_2)$ :- $(solve(G_1) \land solve(G_2)),$} \\ 
\>{\tt $solve(G_1 \lor G_2)$ :- $(solve(G_1) \lor solve(G_2)),$} \\
\>{\tt $solve(\somex{x} G)$ :- $(\somex{x} solve(G)),$ {\rm and}} \\
\>{\tt $solve(\allx{x} G)$ :- $(\allx{x} solve(G)),$}
\end{exmple}
\noindent 
and a ``clause'' for the atomic case that results in setting up
argument registers and then calling the appropriate predicate.  The
clauses for $solve$ will themselves be compiled, and this results in a
partial compilation of the actual goal that is produced from $P(a)$ at
run-time.  Note that the clauses for $solve$ do not include one for
the case of a dynamically created implication goal.  The reason for
this is that such a goal will never be produced in the context of our
higher-order language: implications are prohibited from appearing in
(lambda) terms.  This situation is fortunate since it is not clear
that a clause that can be compiled by the methods described in this
paper can be provided for $solve$ for this case.

The higher-order theory of hereditary Harrop formulas accounts for
most of the examples presented in Section \ref{sec:examples}.
However, there is one example that lies outside this theory and this
corresponds to the final definition of {\tt rev}. We reproduce this
definition below (once again using pseudo-Prolog syntax):
\begin{tabbing}
\hspace{2em}\={\tt ($\all$rev\_aux$\,$((}\=\hspace{2em}\=\kill
\> {\tt rev(L1,L2) :-}\\
\> {\tt ($\all$rev\_aux$\,$((rev\_aux([],L2) $\land$}\\
\>\> {\tt ($\all$X$\,\all$L1$\,\all$L3$\,$(rev\_aux([X|L1],L3) :-
rev\_aux(L1,[X|L3]))))} \\
\>\>\> $\supset$ {\tt rev\_aux(L1,[]))).}
\end{tabbing}
\noindent The body of the clause defining {\tt rev} has the form 
$\all${\tt rev\_aux}$\,(F \supset G)$, where $F$ is a formula that
represents the ``clauses'' 
\begin{exmple}
\> {\tt rev\_aux([],L2).}\\
\> {\tt rev\_aux([X|L1],L3) :- rev\_aux(L1,[X|L3]).}
\end{exmple}
\noindent Notice, however, that these formulas are not really program
clauses according to the current definition: the symbol {\tt rev\_aux}
being a predicate variable, the ``heads'' of these formulas, \ie, the
expressions that appear on the right of the implication (or to the
left of {\tt :-}) in them, are not rigid atomic formulas as is
required by our definition of $D$ formulas.

The stipulation that the heads of program clauses be rigid atomic
formulas is motivated by programming considerations. A program clause
is to be thought of as a (partial) definition of a procedure, the name
of the procedure that it defines being the top-level predicate symbol
of its head.  Such an interpretation would obviously not be very
meaningful if this predicate symbol is a variable. The
requirement of rigidity rules out this possibility. However, the
example under consideration shows that this requirement is stronger
than what might be needed. Thus, even though {\tt rev\_aux} is a
variable, it will be replaced by a constant {\em before} the clauses
``defining'' it are added to the program and these clauses will
constitute a meaningful procedure definition subsequent to such a
replacement.  Understanding this situation and noting that there is a
useful paradigm embodied in the definition of {\tt rev} under
scrutiny, it seems worthwhile to extend our language to permit such
definitions.  We do this by enlarging our class of goals to include
formulas of the form $\allx{x} F$ not only when $F$ is a goal but also
when $F$ has the property that replacing all free occurrences of $x$
in it by a constant $c$ produces a goal. We intend, of course, that
this acceptability condition for universally quantified goals be
applied recursively. This intention can be embodied in a recursive
definition, as is done in \cite{EGunter91elp}. We do not do provide such a
definition here, hoping that the intuitive content of the proposed
enrichment is clear. In particular, it should be apparent that the
definition of {\tt rev} that is of interest is a bona fide program
clause in the extended sense just described.

We consider now the additions needed for implementing our language
under this extended definition of goals.  An important requirement
from this perspective is that of a means for establishing the identity
of predicate constants, especially of those predicate constants that
are introduced by the processing of universal quantifiers; such a
scheme will be needed, for instance, in determining access to the
clauses in the program.  As we have already noted, every predicate
constant will be assigned a tag under the present implementation
scheme.  We may, thus, think of an extended name for a predicate
constant that is given by attaching the tag for the constant to it
original name.  The tag for ``global'' constants, \ie, for constants
like {\tt rev} in the clause that appears earlier in this section,
will be uniformly $1$ and, hence, will not add much new information to
the name. The tag value will, on the other hand, be a distinguishing
characteristic of each predicate constant that is introduced in the
course of processing a universal quantifier and that is available in a
given context.  In fact, the original name that is chosen for these
constants may be ignored or considered to be a dummy one like {\em
nil}, and the tag alone may be used in settling questions of identity.

In order to make the proposed naming scheme work, it is necessary to
ensure that the tags associated with predicate constants are available
wherever their names are needed. This is obviously the case for all
global predicate constants.  For a predicate constant that results
from instantiating a universal quantifier, this issue needs to be
considered relative to the variable occurrence that the constant
replaces.  When this variable occurrence is within an argument of an
atomic formula, the machinery already in place ensures the
availability of the tag information at the relevant time. In
particular, the variable whose occurrence is being considered will be
categorized as temporary or permanent and a binding and an associated
tag will be determined for it by the processing of the relevant
quantifier and, if the variable occurrence that is of interest is
embedded within a clause in the antecedent of an implication goal,
transmitted to the point of need by the execution of appropriate {\tt
initialize} instructions.  In the case where the quantified variable
occurs as the head of a goal that is to be invoked, the same
considerations ensure that the tag value of the constant that replaces
it will be known prior to the invocation of the goal.  The only
remaining case is that when the variable occurrence constitutes the
``name'' of a predicate defined by a clause in the antecedent of an
implication goal.  Some changes must be made to existing machinery in
order to ensure that tagging information can be used in the desired
manner in this case.  To see this, let us return to the definition of
{\tt rev}.  When the implication goal in its body is processed, such
as in evaluating the query {\tt rev([1,2,3],L)}, an implication point
record will be created. This implication point record must provide
access to the code for a procedure identified by the ``constant''
introduced for {\tt rev\_aux}. The main component of the name of this
constant is, of course, its tag. However, this tag is known only at
run-time. Hence, it cannot be included directly in the statically
generated code that is associated with the implication point record
and used in determining if the procedure being sought is the one
defined by the clauses whose addition the record corresponds to.

The necessary tag information is, nevertheless, available at the time
the implication point record is created and its use can be
accommodated by making some changes to the compilation of implication
goals. In particular, we think of the name of a predicate ``constant''
that is defined by clauses appearing in the antecedent of an
implication goal as being given by a name and an offset. This offset
is not used in the case of a global predicate constant, and the code
for locating defining clauses for such a constant retains the shape
described earlier.  On the other hand, if the constant is one that is
introduced by processing a universal quantifier, then the name
component, which we will consider to be {\em nil}, becomes irrelevant,
and the offset indicates the location in the environment record that
was current at the time the implication point record was created where
the binding for the quantified variable is stored and from where the
tag may be obtained.  When an attempt is made to solve an atomic goal
or to fill in the vector $nc$ in an implication point record, it may
be necessary to locate code for predicate constants that are
introduced by processing universal quantifiers.  This task is carried
out relative to an implication point record by comparing the tag
associated with the constant and the tags obtained by using the {\tt
E'} field of the record and the offset numbers for the ``hidden''
predicates that are associated with the record.\footnote{We have only
presented a schematic solution to the problem here. In an actual
implementation, the tags for hidden predicates may be precomputed at
the creation of the implication point record and stored in it.
Alternatively, this computation may be carried out when first needed
and stored for subsequent use.}

We consider now the compilation of atomic goals in conjunction with
the scheme outlined above. Atomic goals whose predicate names are
visible at the outermost level are compiled as before by using the
{\tt call} and {\tt execute} instructions. The compilation of an
atomic goal whose name is hidden by an enclosing universal quantifier
requires the use of one of the instructions
\begin{exmple}
\> {\tt call\_value Vi,n}\\
\> {\tt execute\_value Vi} 
\end{exmple}
\noindent where {\tt Vi} is a temporary or permanent variable. These
instructions differ from the {\tt call} and {\tt execute} instructions
only in the way they determine the location of the code to be invoked:
this is done by determining the tag value associated with the constant 
that {\tt Vi} is bound to and, assuming that this value is $t$, then
searching from the most recent implication point record for code named
by $\langle${\em nil}$,t \rangle$.

The code that would be produced for {\tt rev} using the ideas
presented in this section is shown below. We assume in this code that
{\tt t1} is a pointer to the table created for the implication goal
that appears in the body of the clause defining {\tt rev}.

\begin{tabbing}
\quad\={\tt rev\_aux: }\={\tt push\_impl\_point c1,4}\quad\=\kill

\>{\tt rev:}\>{\tt allocate 2}\>{\tt \% rev\_aux, L2 are permanent
variables}\\  
\>\>{\tt get\_variable Y2,A2 }\\
\>\>{\tt incr\_universe}\>{\tt \% ( $\all$} \\
\>\>{\tt set\_univ\_tag Y1}\>{\tt \% \ \ \ \ rev\_aux}\\ 
\>\>{\tt push\_impl\_point   t1,2}\>{\tt \% \ \ \ \ \ \ add rev\_aux
code}\\ 
\>\>{\tt put\_constant [],A2 }\>{\tt }\\
\>\>{\tt call\_value  Y1,2 }\>{\tt \% \ \ \ \ \ \ \ call rev\_aux}\\ 
\>\>{\tt pop\_impl\_point }\>{\tt \% \ \ \ \ \ \  restore earlier
program }\\ 
\>\>{\tt decr\_universe}\>{\tt \% ) }\\
\>\>{\tt deallocate}\>{\tt }\\
\>\>{\tt proceed}\>{\tt }\\
\end{tabbing}

The code that would be generated for the clauses in the antecedent of
the implication goal that appears in the body of the definition of
{\tt rev} is shown below. The label {\tt $\langle$nil,1$\rangle$} is
used here to indicate that this code is indexed by a predicate
constant whose name component is {\em nil} and whose offset is $1$.
\begin{tabbing}
\quad\={\tt rev\_aux: }\={\tt push\_impl\_point c1,4}\quad\=\kill

\>{\tt $\langle$nil,1$\rangle$:}\>{\tt try\_me\_else C1}\>{\tt }\\
\>\>{\tt     initialize X3,2 }\>{\tt \% X3 = L2}\\
\>\>{\tt     get\_constant [],A1}\>\\
\>\>{\tt     get\_value X3,A2 }\>{\tt \% unify L2 and second
argument}\\ 
\>\>{\tt     proceed}\>{\tt }\\
\\
\>{\tt C1:}\>{\tt trust\_me\_else fail}\>{\tt }\\
\>\>{\tt     initialize X3,1}\>{\tt \% X3 = rev\_aux}\\
\>\>{\tt     get\_list A1  }\>{\tt }\\
\>\>{\tt     unify\_variable X4}\>{\tt }\\
\>\>{\tt     unify\_variable A1}\>{\tt }\\
\>\>{\tt     get\_variable X5,A2}\>{\tt }\\

\>\>{\tt     put\_list  A2}\>{\tt }\\
\>\>{\tt     set\_value X4}\>{\tt }\\
\>\>{\tt     set\_local\_value X5}\>{\tt }\\
\>\>{\tt     execute\_value X3}\>{\tt }\\
\end{tabbing}
\smallskip

\noindent 
This code should be compared with the code shown for {\tt rev\_aux} in
Section~\ref{sec:compile}. The scoping effect of the universal
quantifier warrants the conclusion that the second clause for {\tt
rev\_aux} is the last one that can be used for solving it and,
consequently, that the choice point record can be discarded prior to
using it.

The scoping effect of the universal quantifier actually permits
further improvements to be made to the code shown above for {\tt rev}
and {\tt rev\_aux}. First, the location of code that must be used in
solving the consequent of the implication goal in the body of the
clause for {\tt rev} can be determined statically to be that which is
labelled with {\tt $\langle$nil,1$\rangle$}. The {\tt call\_value}
instruction that appears in the code for {\tt rev} can, therefore, be
replaced by a direct call reminiscent of the WAM. A similar
observation applies to the body of the second clause defining {\tt
rev\_aux}, permitting the {\tt execute\_value} instruction appearing
in the code for this predicate to be replaced by an {\tt execute}
instruction like that of the WAM. A further observation is that the
code for {\tt rev\_aux} can be invoked from only these two places and
so the implication point record that would be created in the course of
solving the implication goal in the body of {\tt rev} will not be
needed for the purpose of accessing this code. As already noted, the
two clauses for {\tt rev\_aux} could not be extending a previously
existing definition for this predicate. Thus, the only purpose for the
mentioned implication point record is that it maintains a binding for
the variable {\tt L2} that is free in the first clause for {\tt
rev\_aux}. If an alternative means is provided for remembering this
binding, the creation and removal of the implication point record can
also be dispensed with.

Observations such as those above can lead to significant efficiency
improvements in the code that is produced. It seems worthwhile,
therefore, to develop methods of static analysis that allow such
observations to be made.  Note, however, that, even after such a
static analysis, a complete treatment of the current language will
still require the issues examined in this section to be dealt with. In
particular, there are situations in which definitions of predicates
whose names are hidden by universal quantifiers actually change in the
course of computation. The location of the code for such
predicates can therefore not always be determined statically and some
mechanism must be provided both for dynamically extending existing
definitions and for identifying the relevant code at run-time.  To
understand these comments, let us consider the following goal
(presented, again, in pseudo-Prolog syntax) in which {\tt a} and {\tt
b} are constants and {\tt r} and {\tt s} are predicates that are
defined by clauses in the (implicit) global program:

\begin{tabbing}
\hspace{2em}\={\tt $\all$p$\,\all$q$\,$((}\=\hspace{5em}\=\kill
\> {\tt $\all$p$\,\all$q$\,$((($\all$X$\,$(p(X) :- q(X))) $\land$}\\ 
\>\>{\tt ($\all$Y$\,$(q(Y) :- r(Y))))}\\
\>\>\>{\tt $\supset$ (p(a) $\land$ (($\all$Z$\,$(q(Z) :- s(Z)))
$\supset$ p(b)))).}
\end{tabbing}

\noindent Assume that the constants introduced in processing
the two outermost universal quantifiers are named {\tt p} and {\tt q}
respectively. Then, solving the given goal eventually requires the two
goals {\tt p(a)} and {\tt p(b)} to be solved. The definition of {\tt
p} in both cases is given by the clause
\begin{exmple}
\> {\tt $\all$X$\,$(p(X) :- q(X)).}
\end{exmple}
\noindent Notice, however, that the definition of {\tt q} is different
in the two cases. When {\tt p(a)} is to be solved, {\tt q} will be
defined by the 
sole clause
\begin{exmple}
\> {\tt $\all$Y$\,$(q(Y) :- r(Y)).}
\end{exmple}
\noindent Prior to solving {\tt p(b)}, this definition will be
extended by the addition of the clause 
\begin{exmple}
\> {\tt $\all$Z$\,$(q(Z) :- r(Z)).}
\end{exmple}
\noindent Thus, despite the universal quantification over {\tt q}, the
occurrence of {\tt q} in the clause defining {\tt p} cannot be
compiled into a direct call. Some mechanism that supports the
extension of definitions even for such predicates and that facilitates
the resolution of identity questions pertaining to them therefore
appears necessary.

\section{CONCLUSION}\label{sec:conclusion}

In this paper we have considered an enrichment to logic programming
that is based on allowing implications and universal quantifiers to
appear in goals. We have argued that the inclusion of these symbols
leads to several novel features at a programming level, including a
means for giving names and programs a scope. We have then discussed
the implementation problems that arise from the addition of 
these symbols. These problems are of three broad kinds:

\begin{romanlist}
\item The possibility for existential and universal quantifiers
to occur in mixed order in goals requires a careful treatment 
of unification. In particular, instantiations for variables
must respect the order in which quantifiers appear.

\item Programs may change in the course of computation by the
addition or removal of clauses and a mechanism is needed for
implementing these changes in an incremental fashion. Furthermore,
backtracking may cause a return to a previously existing program
context and so it should be possible to resurrect such contexts
quickly. 

\item A method is needed for representing program clauses that
permits compilation and the sharing of compiled code even though the 
exact form of these clauses may be dynamically determined. 
\end{romanlist}

We have presented solutions to these problems.  Our solution to the
first problem is based on an association of tags with constants and
variables and the use of these tags to ensure that variable bindings
determined during unification respect the necessary constraints.  With
regard to the second problem, we have proposed a new kind of record
called an implication point record that represents the creation of a
new program by the addition of a certain set of clauses to a
previously existing program.  Implication point records are to be
maintained on the local stack and each of them will be retained as
long as there is a possibility to return to the program context that
it represents.  The resurrection of an earlier program context can
therefore be achieved simply by switching to the appropriate
implication point record.  Finally, as a solution to the last problem,
we have described a closure-based representation of program clauses.
This representation separates each clause into a fixed part that can
be compiled (and shared) and an environment that records the part that
is dynamically determined.  A feature of the solutions that we have
developed to the problems described above is that they can all be
easily integrated into the structure of the WAM.  We have described
this integration and have discussed the issue of compiling programs in
our extended language into instructions that will run on the resulting
machine.

Although the focus in this paper has been on a first-order language,
the ultimate objective of our work is to provide an implementation of
a polymorphically typed, higher-order version of this language.  The
ideas that we have developed here for implementing the scoping
mechanisms are, as we have indicated, not dependent on whether these
mechanisms are being added to a first-order or a higher-order
language. There are, however, substantial additional issues that have
to be considered in implementing the desired form of typing and in
realizing higher-order aspects.  We have considered these issues in
detail elsewhere \cite{KNW92,Nad94a,NJW92,NW93a}. We have also
combined the ideas that we have developed for implementing these
aspects with those in this paper to produce an abstract machine for
the overall language.  The development of an emulator for this machine
and of a compiler for translating programs in the extended language
into instructions that will run on this machine is currently being
undertaken.

\section{ACKNOWLEDGEMENTS}

This paper has benefitted from the suggestions for improvements made
by its referees. Some of the ideas in this paper were previously
presented in \cite{JN91} and we are grateful to Evelina Lamma, Dale
Miller and Paola Mello for their comments on this presentation.

\end{document}